\documentclass{aa}  
\usepackage{graphicx}
\usepackage{txfonts}
\usepackage{color}

\newcommand{\ch}[1]{{#1}} 
\newcommand{\chm}[1]{{#1}} 

\usepackage{hyperref}
\hypersetup{
  colorlinks = true,        
  linkcolor = blue,         
  citecolor = cyan,         
}

\graphicspath{{./}}
\bibpunct{(}{)}{;}{a}{}{,} 

\newcommand{\ie}{i.e.,~}
\newcommand{\eg}{e.g.,~}
\newcommand{\cf}{cf.~}

\newcommand{\harm}{\texttt{HARM}~}

\begin{document} 

   \title{Constrained transport and adaptive mesh refinement in the \\ Black Hole Accretion Code}


   \author{Hector Olivares\inst{\ref{inst1}}
          \and
          Oliver Porth\inst{\ref{inst1},\ref{inst2}}
          \and
          Jordy Davelaar\inst{\ref{inst3}}
          \and
          Elias R. Most\inst{\ref{inst1}}
          \and \\
          Christian M. Fromm\inst{\ref{inst1},\ref{inst4}}
          \and
          Yosuke Mizuno\inst{\ref{inst1}}
          \and
          Ziri Younsi\inst{\ref{inst5}}
          \and
          Luciano Rezzolla\inst{\ref{inst1}}
          }

   \institute{Institute for Theoretical Physics, Goethe-University, D-60438,
   Frankfurt am Main, Germany \label{inst1}
   \and
   Astronomical Institute Anton Pannekoek, University of Amsterdam,
   Science Park 904, 1098 XH, Amsterdam, The Netherlands \label{inst2}
   \and
   Department of Astrophysics/IMAPP, Radboud University Nijmegen P.O. Box 9010,
   6500 GL Nijmegen, The Netherlands \label{inst3}
   \and
   Max-Planck-Institut f\"ur Radioastronomie, Auf dem H\"ugel 69,
   D-53121 Bonn, Germany \label{inst4}
   \and
   Mullard Space Science Laboratory, University College London,
   Holmbury St.\,Mary, Dorking, Surrey RH5 6NT, UK \label{inst5}
   }

   \date{\today}

  \abstract
  {Worldwide very-long-baseline radio interferometry (VLBI) arrays are
    expected to obtain horizon-scale images of supermassive black hole
    candidates as well as of relativistic jets in several nearby active
    galactic nuclei. This, together with the expected detection of electromagnetic
    counterparts of gravitational-wave signals, motivates the
    development of models for magnetohydrodynamic flows in strong
    gravitational fields.}
  %
  %
  {The Black Hole Accretion Code (\texttt{BHAC}) is a publicliy available
    code intended to aid with the modelling of such sources by performing
    general relativistic magnetohydrodynamical (GRMHD) simulations in
    arbitrary stationary spacetimes. 
    New additions to the code are required in order to guarantee an accurate  
    evolution of the magnetic field when small and large scales are captured
    simultaneously.  
    }
  %
  %
  {We discuss the adaptive mesh refinement (AMR) techniques employed in
    \texttt{BHAC}, essential to keep several problems computationally
    tractable, as well as staggered-mesh-based constrained transport (CT)
    algorithms to preserve the divergence-free constraint of the magnetic
    field, including a general class of prolongation operators for
    face-allocated variables compatible with them.
    }  
  %
  %
  {After presenting several standard tests for the new implementation, we
    show that the choice of divergence-control method employed can
    produce qualitative differences in the simulation results for
    scientifically relevant accretion problems. We demonstrate the
    ability of AMR to decrease the computational costs of black-hole
    accretion simulations while sufficiently resolving turbulence arising
    from the magnetorotational instability. In particular, we describe a
    simulation of an accreting Kerr black hole in Cartesian coordinates
    using AMR to follow the propagation of a relativistic jet while
    self-consistently including the jet engine, a problem set up-for
    which the new AMR implementation is particularly advantageous.}
  %
  %
  {The CT methods and AMR strategies discussed here are being currently employed
  in the simulations performed with \texttt{BHAC} used in the generation of theoretical
  models for the Event Horizon Telescope Collaboration.
  }


   \maketitle
%


\section{Introduction}

The prospects of horizon-scale images of the two nearest supermassive
black hole (SMBH) candidates Sgr~A* and M~87, soon to be obtained by
very-long-baseline interferometry (VLBI) arrays
\citep{Doeleman2008,Akiyama2015, Fish2016,Goddi2017,Issaoun2019}, open the
possibility to study in great detail both fundamental and astrophysical
aspects of these objects. Among the most exciting possibilities are
direct observations of the black-hole shadow
\citep{Abdujabbarov2015,Younsi2016,Psaltis2016,Psaltis2014,Mizuno2018} or
the movement of hot spots in the accretion flow \citep{Broderick2006}, as
well as deciphering the cause of flares and non-thermal emission
mechanisms \citep{Ozel2000,Dexter2012,Davelaar2018,Davelaar2019}.
In addition to Sgr~A* and M~87, VLBI observations will provide
high-resolution images of other sources of great interest. For instance,
observations of the two-sided jet in the nearby
active galactic nucleus (AGN) Cen~A \citep{Kim2018}
offer unique possibilities to study the the dynamics of jet formation
and propagation in these objects. VLBI images could also be crucial to
discriminate between models for the periodic variability in the BL Lac
object OJ287, namely a secondary SMBH plowing through the accretion disk
\citep{Valtonen2008}, or a precessing disk \citep{Katz1997} or jet
\citep{Britzen2018}.

General-relativistic magnetohydrodynamical (GRMHD) simulations are an
invaluable tool and possibly the only available one to assess the
validity of theoretical models with respect to astronomical observations;
however, an important challenge for codes performing such simulations is
the interplay between different scales relevant for the accretion
process. Such a large excursion of lengthscales can easily make some
problems prohibitive even for massively parallel numerical
simulations. In fact, to aid with the interpretation of astronomical
observations, simulations of accretion flows must reproduce global
quantities of the system, such as accretion rates, spectra and light
curves, or large-scale features such as the position of re-collimation
shocks in the jet. However, many of these observables are determined by
turbulent phenomena occurring at much smaller scales, which must be
resolved to a reasonable degree to properly reproduce the physics. For
instance, the mechanism of angular momentum transport that permits
accretion \citep{Shakura1973} is now believed to be magnetic
turbulence driven by the the magnetorotational instability
\citep[MRI, ][]{Balbus1991}, and the processes re-collimating the jet
and leading to equipartition \citep{Porth2015}, are driven by several
magnetohydrodynamical instabilities \citep[see \eg][]{Mizuno2012}
which interact with the acceleration processes 
\citep[see \eg][]{Aloy:2006rd}.

By concentrating resolution only at places where it is most needed,
Adaptive Mesh Refinement (AMR) offers a flexible solution to this
problem. In addition, a great advantage of AMR is the possibility of the
grid to follow moving features while applying sufficient resolution.
This can be of special importance, for instance, for systems with complex
geometries, such as precessing jets \citep{Britzen2018,Liska2018} and
discs \citep{Fragile2007b,McKinney2013} or tidal disruption events
\citep{Tchekhovskoy2014}.

Among the variety of GRMHD codes reported on the literature
\citep{Hawley84a, Koide00, DeVilliers03a, Gammie03, Baiotti04,
  Duez05MHD0, Anninos05c, Anton05, Mizuno06, DelZanna2007,
  Giacomazzo:2007ti, Radice2012a, Radice2013b, McKinney2014, Etienne2015,
  Zanotti2015, White2016, Meliani2017, Anninos2017, Liska2018,
  Fambri2018} an increasing number is making use of AMR techniques,
\citep[see \eg][]{Anninos05c,Zanotti2015b,White2016,Liska2018}. One of
them is the publicly available \texttt{BHAC} (the {\it Black Hole Accretion
    Code}, \url{www.bhac.science}), which was described in
\citet{Porth2017} and is the main focus of this work.
The consistency of results obtained by several of the codes
in this list is thoroughly analyzed in the forthcoming work of
 \citet{PorthChatterjeeEtAl2019}.

An important challenge to the application of AMR in GRMHD simulations
comes from the solenoidal constraint of the magnetic field,
$\nabla\cdot\boldsymbol{B} = 0$, which must be fulfilled in order for the
simulation to represent a physical state. The numerous schemes devised to
enforce this constraint present different degrees of complication to be
coupled with AMR techniques. Divergence-cleaning schemes, which require
only the solution of an additional equation damping and propagating away
violations to the constraint, can be coupled straightforwardly to AMR
grids, as done by \citet{Anninos05c,Zanotti2015b,Porth2017}.

Unfortunately, comparisons among several schemes in non-relativistic
magnetohydrodynamics \citep{Toth2000,Balsara2004,Mocz2016} show that all
tested variants of the divergence-cleaning method produce important spurious
oscillations in the magnetic-field energy. \citet{Balsara2004} have
attributed these oscillations to the non-locality introduced by the
constraint-damping equation, thus suggesting that all divergence-cleaning
methods may suffer from the same problem. In a previous work
\citep{Porth2017}, we have observed the same behaviour in GRMHD when
comparing simulations on uniform grids performed using the
divergence-cleaning technique known as Generalized Lagrange Multipliers
\citep[GLM,][]{Dedner:2002} and the method known as flux-interpolated
Constrained Transport \citep[flux-CT,][]{Toth2000}.

By adopting a discretisation of Faraday's law consistent with that of the
constraint, Constrained-Transport (CT) methods maintain the solenoidal
condition to machine precision throughout the solution of the GRMHD
equations. In general, this requires to define the components of the
magnetic field at cell faces, in a grid that is staggered with respect to
that of the hydrodynamic variables.

A drawback of these methods is that the fulfilment of the constraint at
every next step depends on its fulfilment at the current step; therefore,
the initial condition must be divergence-free and care must be taken in
order not to generate magnetic-field divergence by any other means. This can
happen, in particular, at boundary cells, at coarse/fine interfaces in
AMR simulations, and when creating and destroying blocks at prolongation
(refining) and restriction (coarsening). Although the generation
divergence at resolution jumps can by avoided by advancing in time the
magnetic vector potential and computing the magnetic field as its curl
\citep{Etienne:2010ui}, the method is sensitive to the gauge condition
employed, giving especially bad results for gauges with zero-speed modes
\citep{Etienne2012a}.

Furthermore, even though cell-centred versions of these methods exist
\citep{Toth2000} and have been applied in GRMHD codes \citep[see
  \eg][]{Gammie03,Mizuno06,Porth2017}, these are incompatible with AMR,
since the problem of finding divergence-preserving prolongation and
restriction operation for cell-centred magnetic fields is
under-determined, \ie information is lost when abandoning the face
representation in favour of the cell-centred representation
\citep{Toth2002}.

The approach used in this work consists in applying divergence-preserving
restriction and prolongation operators for face-allocated magnetic
fields. Examples of these operators were derived independently by
\citet{Toth2002} and \citet{Balsara2001b}, and have been employed in
codes such as \texttt{BATSRUS} \citep{Toth2012} and \texttt{AstroBEAR}
\citep{Cunninghametal09}.

In addition, staggered magnetic fields allow the use of upwind magnetic
field evolution schemes. In fact, as shown by \citet{Flock2010},
not-upwind algorithms such as the widely used flux-CT and the
\citet{Balsara99} method (BS) are prone to numerical instabilities.
Upwind CT schemes include the second-order algorithm by
\citet{Gardiner2005}, used for example in \texttt{Athena++}
\citep{White2016}, and the two algorithms presented in
\citet{londrillo2004divergence} and in \citet{DelZanna2007}, which allow
the use of high-order schemes for the integration of Faraday's law.

This work largely focuses on the AMR-compatible implementation of
staggered grids and the upwind CT methods by
\citet{londrillo2004divergence} and \citet{DelZanna2007} in
\texttt{BHAC}. Special attention is given to a new derivation of
divergence-preserving prolongation operators, which constitute a
coordinate-independent generalisation of those found in the literature
for the special case of Cartesian geometry \citep{Toth2002}.

Though these methods are fully general, technical details on their
implementation in \texttt{BHAC} are provided, with the purpose of
documenting how they can be incorporated in a GRMHD code. These include
the ghost-cell exchange, the data structures and the special treatments
employed at coarse/fine interfaces, and at the polar axis. The new
methods implemented in the code are validated through several tests,
which are also used to highlight the differences between the newly
implemented schemes and those already present and validated. In
particular, we present simulations of magnetised accretion onto a Kerr
black hole, which take advantage of the newly implemented methods. This
problem is of central importance for the interpretations of future images
to be obtained by the EHT, and constitutes the main scientific
application of the code. As shown later, we find that the divergence
control technique employed significantly affects the dynamics of this
system.

Moreover, the use of AMR makes it possible to simulate this system in
Cartesian coordinates while still resolving the growth of the MRI with
sufficient accuracy to reproduce the same mass-accretion rate and fluxes
through the horizon found in spherical coordinates. This might be useful
to avoid and possibly quantify directional biases introduced by grids in
spherical coordinates, such as those commonly used in simulations of the
accretion process, or eliminate the need of special boundary conditions in the
vicinity of the polar axis.

This work is organised as follows: Section \ref{sec:formulation} presents
the formulation of the equations of GRMHD employed in \texttt{BHAC}, as
well as a brief description of the coordinates and equations of state
available in the code. Section \ref{sec:methods} summarizes the numerical
methods used to solve these equations, namely finite-volume (Section
\ref{sec:finite-volume}) and CT methods (Section \ref{sec:CT}). The
mathematical and technical aspects of \texttt{BHAC}'s AMR-compatible
implementation of staggered grids is contained in Section
\ref{sec:AMR-CT}, with Section \ref{sec:stg-amr} focusing on the new
prolongation operators. Section \ref{sec:tests} presents the problems
used to validate the new methods, while the simulations of accretion onto
black holes using the newly implemented methods are described in Section
\ref{sec:accretion}. Finally, in Section \ref{sec:conclusion} we
summarise and discuss the prospects of future research.

Hereafter, we use Greek symbols for spacetime indices that run from 0 to
3 and Latin symbols for for hypersurface indices, that run form 1 to 3.

\section{Mathematical formulation}
\label{sec:formulation}

\subsection{Equations of GRMHD}

\texttt{BHAC} was initiated as an extension of the \texttt{MPI-AMRVAC}
framework towards GRMHD simulations in 1, 2, and 3 dimensions using
finite-volume methods and a variety of modern numerical methods,
described more in detail in \citet{Porth2017}. It exploits much of
\texttt{MPI-AMRVAC}'s infrastructure for parallelisation and block-based
automated Adaptive Mesh Refinement (AMR), resulting in a potentially
significant saving in computational time and computing
resources. \texttt{BHAC} solves the equations of ideal
magnetohydrodynamics (MHD) in general relativity but also in more generic
spacetimes \citep{Mizuno2018}. The GRMHD equations are expressed as
conservation equations for the rest-mass density, the energy and
momentum, and the homogeneous Maxwell equations
\begin{align}
\begin{split}
\nabla_{\mu} (\rho u^{\mu}) =&\ 0 \,, \\
\nabla_{\mu} T^{{\mu \nu}} =&\ 0 \,,  \\
\nabla_{\mu}\, ^{*}\!F^{\mu\nu} =&\ 0 \,,
\end{split}
\label{eq:ideal_MHD_covariant}
\end{align}
where $\rho$ is the particle number density in the fluid frame, $u^\mu$
is the fluid 4-velocity in the observer frame, $T^{\mu \nu}$ is the
energy-momentum tensor and $^{*}\!F^{\mu\nu}$ is the dual of the Faraday
tensor.

The Faraday tensor $F^{\mu\nu}$ and its dual $^{*}\!F^{\mu\nu}$ are such
that the electric and magnetic fields measured by an observer moving at
4-velocity $U^\mu$ are
\begin{equation}
E^\mu \coloneqq F^{\mu\nu}U_\nu \ \ \text{and} \ \ B^\mu \coloneqq ^{*}\!F^{\mu\nu}U_\nu \,.
\label{eq:Faraday}
\end{equation}
and are related as
$^{*}\!F^{\mu\nu} \coloneqq \frac{1}{2}\epsilon^{\mu\nu\alpha\beta}F_{\alpha\beta}$
where $ \epsilon^{\mu\nu\alpha\beta} $ are the components of the
Levi-Civita tensor. Therefore, in a general frame moving with 4-velocity
$U^\mu$
\begin{align}
F^{\mu\nu} \coloneqq U^\mu E^\nu - U^\nu E^\mu - \epsilon^{\mu\nu\lambda\delta}
U_\lambda B_\delta \,,
\label{eq:genericFaraday}
\end{align}
and
\begin{align}
^{*}\!F^{\mu\nu} \coloneqq U^\mu B^\nu - U^\nu B^\mu + \epsilon^{\mu\nu\lambda\delta} U_\lambda E_\delta\,.
\end{align}

Instead of solving the inhomogeneous Maxwell equations, we close the
system by enforcing the ideal-MHD condition 
\begin{align}
  e^\mu = F^{\mu \nu} u_{\nu} = 0 \,,
  \label{eq:ideal}
\end{align} 
where $e^{\mu}$ denotes the electric field in the fluid frame.
Consequently, the magnetic field in the fluid frame is $ b^{\mu} =
^{*}\!F^{\mu\nu} u_\nu$ and
\begin{align}
  ^{*}\!F^{\mu\nu} = b^\mu u^\nu - b^\nu u^\mu \,.
  \label{eq:fmunu}
\end{align}
Physically, the ideal-MHD condition corresponds to a plasma with an
infinite conductivity and has the important consequence that in this case
the magnetic field is simply advected with the fluid (frozen-flux
theorem).

The energy-momentum tensor $T^{\mu \nu}$ includes both fluid and
electromagnetic contributions. Using equation (\ref{eq:fmunu})
to write its electromagnetic part in terms of $b^\mu$
only \citep{Anile1990}, it reads
\begin{align}
T^{\mu\nu} = \rho h_{\rm tot}u^{\mu}u^{\nu} + p_{\rm tot} g^{\mu \nu}
-b^{\mu}b^{\nu}\,.
\label{eq:tmunub}
\end{align}
where $ h_{\rm tot} \coloneqq h + b^{2}/\rho$ is the total specific enthalpy and
$ p_{\rm tot} = p+b^{2}/2$ is the total pressure, both in the fluid
frame.

The system is closed once an equation of state relating the specific
enthalpy of the fluid with the density and the pressure $h=h(\rho,p)$ is
specified. The equations of state for an ideal gas, a Synge gas and a
polytropic fluid are currently implemented in \texttt{BHAC}
\citep[see][for more details]{Porth2017}, although the implementation of
any additional one is straightforward as long as it is
analytic. Additionally, \texttt{BHAC} can be used together with the microphysics
routines developed in \citet{Most2018b} that can handle
temperature-dependent equations of state and provide a neutrino leakage
scheme \citep{Ruffert96b, Oconnor10,Galeazzi2013}.

In order to formulate system \eqref{eq:ideal_MHD_covariant} as a set of
evolution equations, we employ the 3+1 decomposition of the spacetime
\citep[see, \eg][]{Alcubierre:2008,Rezzolla_book:2013}.
The metric is decomposed as $g_{\mu\nu}=\gamma_{\mu\nu} - n_\mu n_\nu$,
where $n_\mu$ and $\gamma_{\mu\nu}$ are such that $\gamma_{\mu}\,^\nu
n_\nu=0$ and $n_\nu n^\nu = -1$. Thus, $\gamma_{\mu}\,^\nu$ defines a
projection operator onto a hypersurface normal to the vector field
$n^\nu$ and contains the 3-metric induced in the hypersurface.

The explicit form of the line element in a 3+1 split spacetime is
\begin{equation}
ds^2 = -\alpha^2 dt^2 + \gamma_{ij}(dx^i+\beta^i dt)(dx^j+\beta^j dt) \,,
\end{equation}
where $\alpha$ and $\beta^i$ are the {\it lapse} and the {\it shift
  vector}, respectively. Using $\gamma_{\mu}\,^\nu$ and $n_\nu$, it is
possible to define variables suitable to be evolved. These are the {\it
  conserved variables} in the Eulerian frame: the number density $D \coloneqq
-\rho u^\nu n_\nu$, the covariant 3-momentum $S_i\coloneqq
n_\mu\gamma_{\nu i} T^{\mu\nu}$, the total energy $U\coloneqq n_\mu n_\nu
T^{\mu\nu}$ and the spatial stress tensor $W^{ij}\coloneqq \gamma_{\mu
  i}\gamma_{\nu j}T^{\mu\nu}$.

In terms of these variables, system (\ref{eq:ideal_MHD_covariant}) can be
written as a set of conservation equations
\begin{equation}
\partial_t (\sqrt{\gamma} \boldsymbol{U}) +
\partial_i (\sqrt{\gamma} \boldsymbol{F}^i) = \sqrt{\gamma} \boldsymbol{S}\,,
\label{eq:conservation}
\end{equation}
in addition to the solenoidal constraint for the magnetic field
$\partial_i \sqrt{\gamma} B^i = 0$. Here, $\gamma$ is the determinant of
the induced three-metric, and the vectors of conserved quantities
$\boldsymbol{U}$, the fluxes $\boldsymbol{F}^i$, and the sources
$\boldsymbol{S}$ are given by
\begin{align}
\begin{split}
\boldsymbol{U} = 
\left[
\begin{array}{c}
D  \\
S_{j}  \\
\tau \\
B^{j}
\end{array}
\right] \,, \ \qquad 
\boldsymbol{F}^{i} = 
\left[
\begin{array}{c}
\mathcal{V}^{i} D \\
\alpha W^{i}_{j} - \beta^{i} S_{j} \\
\alpha (S^{i}-v^{i} D) - \beta^{i} \tau \\
\mathcal{V}^{i}B^{j} - B^{i}\mathcal{V}^{j}
\end{array}
\right] \,, \\
\boldsymbol{S} = 
\left[
\begin{array}{c}
0  \\
\frac{1}{2}\alpha W^{ik}\partial_{j}\gamma_{ik} + S_{i}\partial_{j}\beta^{i} - U\partial_{j}\alpha \\
\frac{1}{2} W^{ik} \beta^{j} \partial_{j} \gamma_{ik} + W_{i}^{j}\partial_{j}\beta^{i} - S^{j} \partial_{j} \alpha \\ 
0
\end{array}
\right] \,,\label{eq:vectors}
\end{split}
\end{align}
where $v^i\coloneqq\gamma^{i}_\mu u^\mu/\Gamma$ is the 3-velocity,
$\Gamma \coloneqq -u^\mu n_\mu$ is the Lorentz factor\footnote{This
  quantity is often also indicated as $W$ \citep{Anton05,
    Rezzolla_book:2013}.}, and $\mathcal{V}^{i}\coloneqq\alpha v^{i} - \beta^{i}$
is the {\it transport velocity}. Evolving $\tau\coloneqq U-D$ instead of
$U$ makes the evolution more accurate in regions of low energy and allows
to recover the Newtonian limit.

To calculate the fluxes $\boldsymbol{F}^i$, knowledge of the primitive
variables $\boldsymbol{P}=\left[\rho, \Gamma v^i, p, B^i \right]$ is
required. While it is straightforward to find
$\boldsymbol{U}(\boldsymbol{P})$, $\boldsymbol{P}(\boldsymbol{U})$
requires numerical inversion. The inversion process then consists of
finding the auxiliary variables $\boldsymbol{A}=\left[\Gamma, \xi
  \right]$, where $\xi\coloneqq\Gamma^2\rho h$ and $h$ is the specific
enthalpy compatible with $\boldsymbol{U}$ and $\boldsymbol{P}$, which is
in turn used to find the primitives. Once $\boldsymbol{A}$ is found,
\texttt{BHAC} stores it in order to facilitate new inversions, thus
extending the array $\boldsymbol{U}(\boldsymbol{P})$, as detailed in
\citet{Porth2017}.

\subsection{Coordinates in \texttt{BHAC}}
\label{sec:coordinates}

\begin{table*}[h]
\caption{{\bf Coordinates available in \texttt{BHAC}: } The first column shows
  the identifier of the coordinates within the code, the second their
  name, the third whether the metric derivatives are calculated
  numerically, and the fourth whether they are initialised from the
  4-metric $g_{\mu\nu}$.}
\begin{center}
\begin{tabular}{llll}
\hline
Identifier & Coordinates & Num. derivatives & Init. $g_{\mu\nu}$ \\
\hline
{\tt bl} & Boyer-Lindquist &  No & No \\
{\tt cart} & Cartesian & No & No\\
{\tt cks} & Cartesian Kerr-Schild (CKS) & Yes & Yes \\
{\tt cmks} & Cylindrified modified Kerr-Schild & Yes & No \\
{\tt dleh} & Non-rotating dilaton black hole \citep{Garcia1995} & No & No \\
{\tt ht} & Hartle-Thorne \citep{Hartle68} & Yes  & Yes \\
{\tt ks} & Kerr-Schild & No & No \\
{\tt lrzks} & Horizon penetrating Rezzolla \& Zhidenko coordinates & Yes  & Yes\\
{\tt mks} & Modified Kerr-Schild \citep[MKS,][]{McKinney2004} & No & No \\
{\tt num} & Numerical & Yes & Yes/No \\
{\tt rns} & RNS \citep{Stergioulas95} & Yes & No \\
{\tt rz} & Rezzolla \& Zhidenko parametrization \citep{Rezzolla2014} & Yes  & No\\
{\tt sp} & Spherical coordinates & No  & No\\
{\tt ss} & Schwarzschild coordinates & No  & No\\
\hline
\end{tabular}
\end{center}
\label{tab:coordinates}
\end{table*}
The main target application of \texttt{BHAC} is the simulation of
accretion onto compact objects in arbitrary spacetimes. This has
allowed to simulate accretion onto Kerr black holes to build
theoretical models of M~87 \citep{Davelaar2019}, but also to
explore the consequences of alternative theories of
gravity \citep{Mizuno2018} or of the presence of a boson star
\citep{Olivares2018} at the Galactic Center in the forthcoming
horizon-scale VLBI images of Sgr A*, as well as the study of
quasi-periodic oscillations in accretion discs around neutron stars
\citep{deAvellar2017}.

As a result, \texttt{BHAC} is designed with a great flexibility to adopt
new spacetime metrics. Its modular structure not only allows us to add
straightforwardly new analytic coordinates, but another interesting
feature is that it is not compulsory to provide derivatives of the
metric, which are used to calculate the sources. The user needs only to
provide the metric functions, and derivatives can be calculated
internally using the complex-step derivative \citep{Squire1998}, a highly
accurate numerical method which is capable of achieving machine precision
for derivatives of algebraic functions \citep{Martins2003}.

The code can handle spacetime metrics given either in 3+1 form (\ie
expressed in terms of $\gamma_{ij}$, $\beta_i$ and $\alpha$) or as the
full 4-metric $g_{\mu\nu}$, as the necessary conversions to 3+1 form are
done internally. In addition, it is able to read tabulated numerical
metrics when the spacetime is known only numerically. A list of metrics
currently available in the code is reported in Table
\ref{tab:coordinates}, which complements that presented in
\citet{Porth2017}.

The need to store the metric components both at barycentre positions and
at cell interfaces is handled efficiently by adopting the metric data
structure described in \citet{Porth2017}, which takes advantage of the
symmetry of the metric components and the fact that the metric and its
derivatives are usually sparsely populated, \ie many of the tensor
components are zero.

\subsubsection{Modified Kerr-Schild coordinates}

Due to its use in later parts of this work, we next describe the less
well known {\it modified Kerr-Schild coordinates} (MKS). MKS coordinates
$x^\mu=(t_{\rm MKS},s,\vartheta,\phi_{\rm MKS})$ were introduced by
\citet{McKinney04} in order to simulate large domains and, at the same
time, concentrate the resolution near the black hole and on the equatorial
plane. They are related to the standard Kerr-Schild coordinates
$x^{\mu'}=(t_{\rm KS},r_{\rm KS},\theta_{\rm KS},\phi_{\rm KS})$
by the transformation
\begin{align}
\begin{split}
t_{\rm KS}&=\ t_{\rm MKS}\,,\\
r_{\rm KS}&=\ R_0 + e^s\,,\\\
\theta_{\rm KS}&=\ \vartheta + \frac{\chm{\vartheta_0}}{2}\sin( 2\vartheta) \,,\\
\phi_{\rm KS}&=\ \phi_{\rm MKS}\,,
\end{split}
\label{eq:mks}
\end{align}
where $R_0$ and $\chm{\vartheta_0}$ are constant parameters. Note that the convention
for the parameter $\chm{\vartheta_0}$ is different from that used in \citet{McKinney04}.
The one employed here is chosen so that $\vartheta$ reduces to
$\theta_{\rm KS}$ when $\chm{\vartheta_0}=0$, thus retrieving logarithmic Kerr-Schild
coordinates.

It is worth to mention that, in contrast to \citet{Porth2017}, in this
work the inverse relation $\vartheta(\theta_{\rm KS})$ is not
approximated using a cubic polynomial, but the resulting transcendental
equation is solved numerically whenever necessary.

\section{Numerical methods}
\label{sec:methods}

\subsection{Finite-volume scheme}
\label{sec:finite-volume}

To introduce the notation used in this work, we next summarise the
finite-volume scheme used by \texttt{BHAC} to evolve the hydrodynamic
variables in system (\ref{eq:conservation}). This scheme was already
described in \citet{Porth2017}, to which we refer the reader for more
details.

To obtain the discretized equations, the simulation domain is divided
into control volumes and the system is integrated over each of them. This
leads to evolution equations for the average
$\boldsymbol{\bar{U}}_{i,j,k}$ of the conserved quantities inside each
cell
\begin{align}
\begin{split}
\frac{d\boldsymbol{\bar{U}}_{i,j,k}}{dt} = 
- \frac{1}{\Delta V_{i,j,k}}
\Biggl[
&\boldsymbol{F^{1}}\Delta S^{1}\bigr|_{i+1/2,j,k} - \boldsymbol{F^{1}}\Delta S^{1}\bigr|_{i-1/2,j,k} + \\
&\boldsymbol{F^{2}}\Delta S^{2}\bigr|_{i,j+1/2,k} - \boldsymbol{F^{2}}\Delta S^{2}\bigr|_{i,j-1/2,k} + \\
&\boldsymbol{F^{3}}\Delta S^{3}\bigr|_{i,j,k+1/2} - \boldsymbol{F^{3}}\Delta S^{3}\bigr|_{i,j,k-1/2} \Biggr]+ \boldsymbol{\bar{S}}_{i,j,k} \,.
\label{eq:fvolume}
\end{split}
\end{align}
Quantities indicated as, say, $\boldsymbol{F^{1}}\Delta
S^{1}\bigr|_{i+1/2,j,k}$ represent integrals of the fluxes over the
surfaces $\Delta S^{1}\bigr|_{i+1/2,j,k}$ bounding the control volume,
and $\boldsymbol{\bar{S}}_{i,j,k}$ is the volume average of the sources.
Both kinds of integrals are approximated to second order, by assigning to
$\boldsymbol{F^{n}}$ ($\boldsymbol{n}=1,2,3$) the point value of the flux
at the face center and to $\boldsymbol{\bar{S}}_{i,j,k}$
the point value at the cell barycentre, \ie at

\begin{align*}
\begin{split}
x^i_{({\rm face}\ k)} &= x^i + \delta^i_k \frac{\Delta x^i}{2} \,,\\
x^i_{({\rm barycentre})} &= \frac{\int_{\rm cell} \sqrt{\gamma}\ x^i \ dx^1 dx^2 dx^3}{\int_{\rm cell} \sqrt{\gamma} \ dx^1 dx^2 dx^3} \,,
\end{split}
\end{align*}

\noindent
respectively, where $\Delta x^i$ is the grid spacing
in each direction and $\delta^i_k$ is Kronecker's delta.
$\boldsymbol{F^{n}}$ is obtained through the approximate
solution of a Riemann problem at the interface, and static integrals such
as cell volumes, interface areas and barycentre positions are calculated
at initialisation using fourth-order Simpson's rule and stored in memory.

The Riemann solvers currently available in \texttt{BHAC} are the Rusanov
method, also known as total variation diminishing Lax-Friedrichs scheme,
and the HLL solver of \citet{Harten83}. The left and right states for the
Riemann problem are obtained using limited reconstructions from the cell
centres \citep[see also][]{Toro09, Rezzolla_book:2013}.

\texttt{BHAC} features a variety of reconstruction schemes, some of which
are listed in \citet{Keppens2012,Porth2017}. They include both
second-order, total variation diminishing (\eg {`minmod'}, {`vanLeer'}),
and high-order such as {`PPM'} \citep{Colella84}, {`LIMO3'}
\citep{Cada2009} and {`MP5'} \citep{suresh_1997_amp}. Recent additions to
that list are the high-order schemes WENO5 and WENOZ+
\citep{Acker2016}. Equation \ref{eq:fvolume} can be integrated using any
of the methods present in the \texttt{MPI-AMRVAC} toolkit. These include
the simple half-step modified Euler, the third order Runge-Kutta RK3
\citep{Gottlieb98} and the strong-stability preserving $s$-step,
$p$th-order RK schemes SSPRK($s$,$p$) schemes: SSPRK(4,3), SSPRK(5,4) due
to \citet{Spiteri2002} \citep[see][for implementation
  details]{Porth2014}.

\subsection{Constrained Transport}
\label{sec:CT}

Constrained transport (CT) is a divergence-control method first proposed
by \citet{Evans1988}. In essence, instead of eliminating the divergence
of the magnetic field once it is created, it modifies the way in which
magnetic-field transport is evolved so as to prevent the creation of
divergence.  CT is able to keep a discretisation of
$\nabla\cdot\boldsymbol{B}=0$ to a precision close to that of floating
point operations by ensuring that the sum of the magnetic fluxes through
the surfaces bounding a cell is zero to machine precision.  Recalling the
definition of the divergence of a vector field $\boldsymbol{B}$
\begin{equation}
\label{eq:divB-definition}
\nabla\cdot\boldsymbol{B}=\lim\limits_{\Delta V \rightarrow 0}
       \frac{1}{\Delta V} \oint_{S=\partial \Delta V} \mkern-27mu
       \boldsymbol{B}\cdot \boldsymbol{n}
       \,dS\,,
\end{equation}
is equivalent to $\nabla\cdot\boldsymbol{B}=0$ in the continuous limit.
When the limit is not taken, \ie for the finite-volume case, it follows
from the divergence theorem that the \textit{average} value of the
divergence $\left(\overline{\nabla\cdot\boldsymbol{B}}\right)_{\rm cell}$ is zero within the
cell.

\begin{figure}[h]
\centering
\includegraphics[width=14pc]{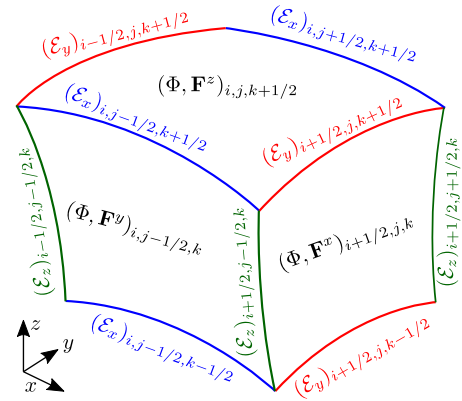}
\caption{\label{fig:CT-algorithm} Spatial location of variables for a
  cell with indices $(i,j,k)$,
  \ch{correspondent to directions $(x,y,z)$, respectively}.
  Line integrals of the electric field
  $\mathcal{E}$ are located at its edges, and magnetic and numerical
  fluxes $\Phi$ and $\boldsymbol{F}^i$ (the latter used for the BS
  algorithm) are located at its faces. The conserved hydrodynamic
  variables belonging to the array $\boldsymbol{U}$, are located
  at cell centres (not shown).}
\end{figure}

The central idea from constrained transport is to give the
electromagnetic variables a special spatial location, as depicted in
Fig. \ref{fig:CT-algorithm}. In particular, to each face of the cell
corresponds a magnetic flux calculated, for example, as
\begin{equation}
\Phi_{i+1/2,j,k}=\int_{\partial V (x^1_{i+1/2})} \gamma^{1/2} B^1
dx^2dx^3 \,,
\end{equation}
and on each edge is associated a line integral of the electric field,
similar to
\begin{equation}
\label{eq:integralG}
\mathcal{E}_{i+1/2,j+1/2,k} = - \int_{x^3_{k-1/2}}^{x^3_{k+1/2}}
\left. E_3\right|_{x^1_{i+1/2},x^2_{j+1/2}} dx^3 \,.
\end{equation}

The magnetic flux at each face is updated from the circulation of the
electric field, using the integral form of Faraday's law
\begin{align}
\label{eq:CT_update}
\frac{d}{dt}\Phi_{i+1/2,j,k} = \phantom{-} &\mathcal{E}_{i+1/2,j+1/2,k}
 - \mathcal{E}_{i+1/2,j-1/2,k}\\ \nonumber
 - \,& \mathcal{E}_{i+1/2,j,k+1/2}
 + \mathcal{E}_{i+1/2,j,k-1/2}\,,
\end{align}
Since each of the line integrals of the electric field is shared by two
faces, but appears with opposite sign in the time update formula for each
of them, the rate of change of $\left(\overline{\nabla\cdot\boldsymbol{B}}\right)_{\rm cell}$,
which can be calculated as the sum of the rate of change of the outgoing
flux through all faces, vanishes. Therefore, the CT time update ensures
that $\left(\overline{\nabla\cdot\boldsymbol{B}}\right)_{\rm cell}$ is kept constant at machine
precision from one iteration to the next.

In order for $\boldsymbol{B}$ to be divergence-free during the whole
simulation, $\left(\overline{\nabla\cdot\boldsymbol{B}}\right)_{\rm cell}$ must be zero at the
initial condition. This can be accomplished by initialising the line
integrals of magnetic vector potential along cell edges and calculating
the initial magnetic fluxes as the circulation of the vector potential
around the cell faces, in the same way as the rate of change of the
magnetic flux is calculated from the circulation of the electric field.

After the time update, the magnetic field is interpolated to the cell
center in order to use it for the inversion from conserved to primitive
variables.

The idea of staggering the magnetic components of the electromagnetic
field to achieve a divergence-free evolution to machine precision was
first proposed by \citet{Yee66}. However, although his method was widely
known in engineering, a staggered grid was not applied in GRMHD until
the work of \citet{Evans1988}.

In the formulas written so far, no approximations have been made.
Approximations come into play when deciding how to calculate the line
integrals of the electric field. The way these approximations are done
distinguishes the different `variants' of CT. Three of these variants are
described in the following sections.

\subsection{Arithmetic averaging (BS)}

This variant was introduced by \citet{Balsara99}. In the comparison
work of \citet{Toth2000}, it is referred to as \textit{Flux-interpolated
  Constrained Transport} (flux-CT), although in most of the literature
this name is used exclusively for the cell-centered version of this
method, proposed also in \citet{Toth2000}. \ch{In this work, we will
use the abbreviation `flux-CT' to refer to the cell-centered version
of the method and `BS' to refer to the staggered one. Both variants are}
particularly suitable for finite-volume schemes, since the electric field
at cell edges is estimated as the arithmetic average of the numerical
fluxes returned by the Riemann solver. In fact, the numerical fluxes
corresponding to the magnetic-field components are surface integrals of
the electric field, for example, the flux in the $x^2$-direction for
$B^1$ is
\begin{equation}
\label{eq:fluxEfield_integrals}
\left. \Delta S^2 \bar{F}^2\right|_{i,j+1/2,k} = \int_{x^1_{i-1/2}}^{x^1_{i+1/2}}\int_{x^3_{k-1/2}}^{x^3_{k+1/2}}\left. E_{x^3}\right|_{j+1/2} \, dx^3 \, dx^1 \,.
\end{equation}
The innermost integral is the same as that of Eq.~(\ref{eq:integralG}),
so the average flux can be interpreted as
\begin{equation}
\left. \Delta S^2 \bar{F}^2 \right|_{i,j+1/2,k} = -  \Delta x_i \, \tilde{\mathcal{E}}_{i,j+1/2,k} \,,
\end{equation}
where $\tilde{\mathcal{E}}_{i,j+1/2,k}$ is the mean value of the integral
from Eq.~(\ref{eq:integralG}) \ch{over the face at $j+1/2$}.
To second-order accuracy, this integral
takes the value $\tilde{\mathcal{E}}_{i,j+1/2,k}$ at the middle of the
cell. Therefore, $\mathcal{E}_{i+1/2,j+1/2,k}$ can be found by
interpolating the averaged fluxes from the four adjacent cell faces as
\begin{align}
\begin{split}
\mathcal{E}_{i+1/2,j+1/2,k} =  \frac{1}{4}
\Biggl(
&\frac{\Delta S^2 \bar{F}^2  \bigr|_{i,j+1/2,k}}{\Delta x_{i}}
+\frac{\Delta S^2 \bar{F}^2 \bigr|_{i+1,j+1/2,k}}{\Delta x_{i+1}} \\
-&\frac{\Delta S^1 \bar{F}^1 \bigr|_{i+1/2,j,k}}{\Delta y_j}
-\frac{\Delta S^1 \bar{F}^1 \bigr|_{i+1/2,j+1,k}}{\Delta y_{j+1}}\Biggr) \,.
\end{split}
\label{eq:interpolationG_fluxes}
\end{align}
Although this algorithm, especially in its cell-centered version, is
widely used in the community \citep[see \eg][]{Gammie03,Noble2009}, it
is known to lack a proper upwinding and to not reduce to the correct
limit for one-dimensional flow \citep{Gardiner2005}, as well as to cause
numerical instabilities \citep{Flock2010}. Efforts to overcome this
problems motivated the development of methods such as those by
\citet{Gardiner2005,londrillo2004divergence,DelZanna2007}. In the next
section, we will briefly describe the methods by
\citet{londrillo2004divergence} and \citet{DelZanna2007}, which we have
also implemented in \texttt{BHAC}.

\subsection{Upwind constrained transport (UCT)}

\begin{figure*}
\includegraphics[width=0.45\linewidth]{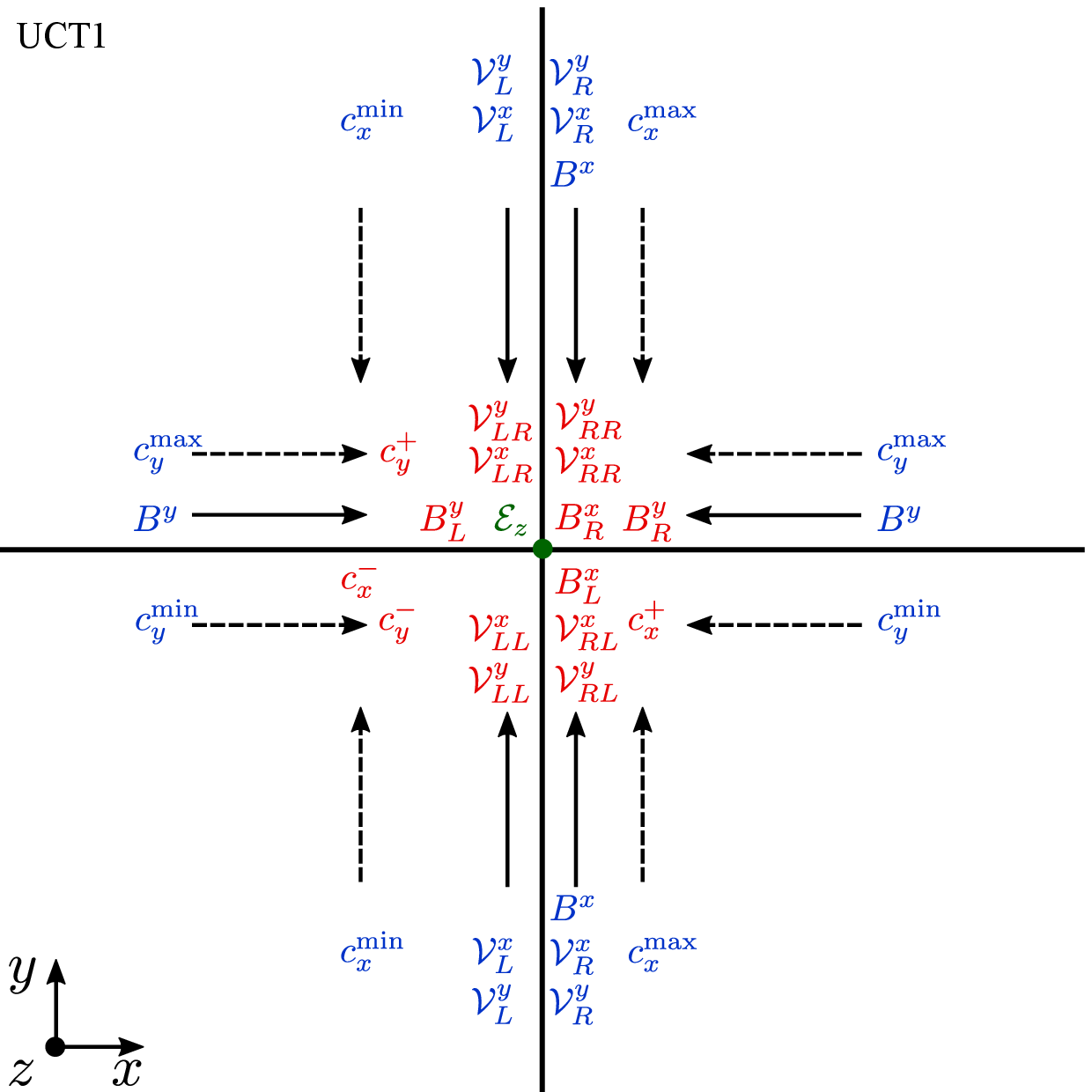}
\hspace{0.1\linewidth}
\includegraphics[width=0.45\linewidth]{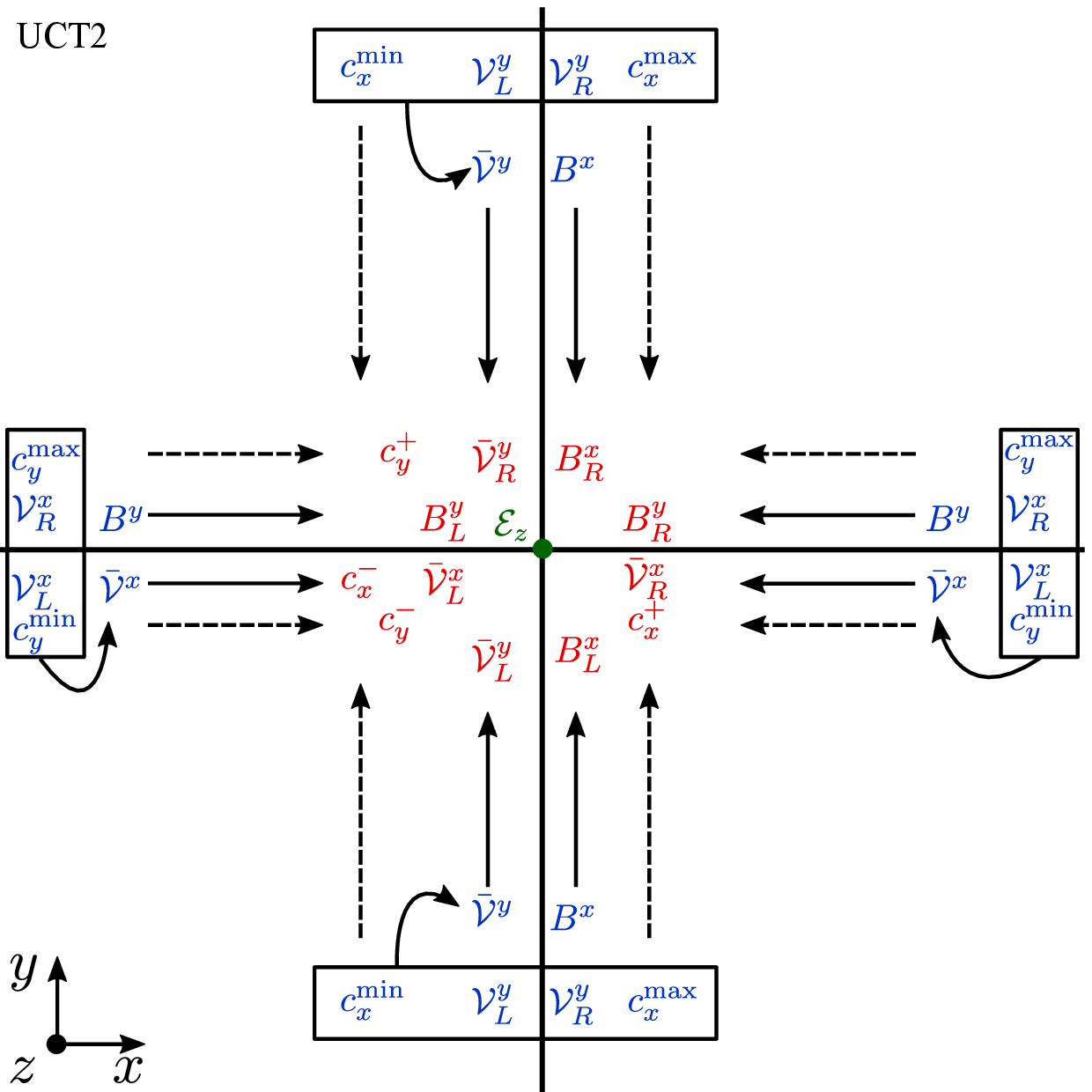}
\caption{Reconstruction of quantities for UCT1 ({\it left}) and
  UCT2 ({\it right}). Quantities in blue are defined at cell faces
  \ch{in $x$ and $y$} and quantities in red at the cell edge \ch{in $z$.
  The resulting line integral of the electric field $\mathcal{E}_z$
  over the edge is shown in green}. Continuous arrows represent
  limited reconstructions \ch{(steps 2 and 3 of UCT1, and 3 of UCT2)} and dashed
  lines the choice of the maximum
  value \ch{(step 5 of UCT1 and 4 of UCT2, the labels $L$ and $R$ for the
  characteristic speeds are omitted in the figure to avoid saturation)}.
  The  average transverse transport velocities \ch{needed} in UCT2
  \ch{(step 2)} $\bar{\mathcal{V}}^{x,y}$ are computed using the
  quantities in the boxes.}
\label{fig:UCT}
\end{figure*}

Upwind Constrained Transport (UCT) is a method to evolve the magnetic
field, first proposed in \citep{londrillo2004divergence}. As a CT method,
like BS, it maintains the divergence of the magnetic field to machine
precision; however, it furthermore aims to accurately reproduce the
magnetic-field continuity and transport properties by using limited
reconstructions. Also, in contrast to arithmetic averaging, UCT reduces
to the correct 1-dimensional limit when symmetry in the other two
directions is assumed. Two variants of UCT are implemented in
\texttt{BHAC}. These are presented in \citep{londrillo2004divergence}
and in \citep{DelZanna2007}, and will be referred, respectively, as UCT1
and UCT2. Now we describe the procedure to obtain $\mathcal{E}$ in each
of these algorithms. To simplify the notation, we rename $(x^1,x^2,x^3)$
as $(x,y,z)$. We focus on the calculation of $\mathcal{E}_z$, but the
other cases can be obtained by iteratively replacing $x\rightarrow y$,
$y\rightarrow z$ and $z\rightarrow x$ (\ch{see Figure \ref{fig:UCT}}). In
\texttt{BHAC}, the implementation of UCT1 is as follows:
\begin{enumerate}
\item At the time of calculating the numerical fluxes at $x$-interfaces,
  store the \ch{characteristic speeds $c^\mathrm{min}_i$ and $c^\mathrm{max}_i$
  ($i=x,y,z$) obtained from the Riemann solver},
  as well as the transport velocity $\mathcal{V}^x_L$,
  $\mathcal{V}^y_L$, $\mathcal{V}^x_R$, and $\mathcal{V}^y_R$.

\item Reconstruct transport velocities along direction $y$ towards the
  edges $z$ and obtain one correspondent to the four states around them
  $\mathcal{V}^i_{LL}$, $\mathcal{V}^i_{LR}$, $\mathcal{V}^i_{RL}$ and
  $\mathcal{V}^i_{RR}$ ($i=x,y$).

\item Reconstruct the magnetic fields on the faces to the same edges and
  obtain $B^x_L$, $B^x_R$, $B^y_L$ and $B^y_R$.

\item Calculate approximations to the electric field on the edge from
  each of the possible states as
  \begin{align*}
E_z^{LL}&=\mathcal{V}^y_{LL} B^x_{L} - \mathcal{V}^x_{LL} B^y_{L}\,, \\
E_z^{LR}&=\mathcal{V}^y_{LR} B^x_{R} - \mathcal{V}^x_{LR} B^y_{L}\,, \\
E_z^{RL}&=\mathcal{V}^y_{RL} B^x_{L} - \mathcal{V}^x_{RL} B^y_{R}\,, \\
E_z^{RR}&=\mathcal{V}^y_{RR} B^x_{R} - \mathcal{V}^x_{RR} B^y_{R}\,.
\end{align*}

\item Take the maximum \ch{characteristic speeds} in each direction from the four
  states,
\begin{align*}
c^{-}_x &= \max{(c^\mathrm{min}_{x,\ L},c^\mathrm{min}_{x,\ R})}\,, \\
c^{+}_x &= \max{(c^\mathrm{max}_{x,\ L},c^\mathrm{max}_{x,\ R})}\,, \\
c^{-}_y &= \max{(c^\mathrm{min}_{y,\ L},c^\mathrm{min}_{y,\ R})}\,, \\
c^{+}_y &= \max{(c^\mathrm{max}_{y,\ L},c^\mathrm{max}_{y,\ R})}\,,
\end{align*}
\ch{where $c^\mathrm{min}_{i,\ L(R)}$ ($c^\mathrm{max}_{i,\ L(R)}$)
are defined as positive in the $-i$ ($+i$) direction, and they
are zero otherwise.}

\item Finally, approximate the electric field, as an average of those at
  the four states weighted by the \ch{characteristic speeds} and corrected by the
  transport of magnetic field, using the formula by
  \citet{londrillo2004divergence}:
\begin{align}
\label{eq:UCT1}
\begin{split}
E_z = \sqrt{\gamma} \left[ \frac{c^{+}_x c^{+}_y E_z^{LL} + c^{+}_x c^{-}_y E_z^{LR}
          + c^{-}_x c^{+}_y E_z^{RL} + c^{-}_x c^{-}_y E_z^{RR}}
          {(c^{+}_x + c^{-}_x)(c^{+}_y + c^{-}_y)} \right. \\
      \left.
      + \frac{c^{+}_x c^{-}_x}{c^{+}_x + c^{-}_x}(B^R_y - B^L_y)
          - \frac{c^{+}_y c^{-}_y}{c^{+}_y + c^{-}_y}(B^R_x - B^L_x) \right] \,.
\end{split}
\end{align}
Since this is a point value, we take the second order approximation that
$\mathcal{E}_z=E_z \Delta z$ and use it to calculate the circulation.
\end{enumerate}

The implementation of UCT2 is as follows:

\begin{enumerate}
\item At the time of calculating the numerical fluxes at each interface,
  store the \ch{characteristic speeds $c^\mathrm{min}_i$ and $c^\mathrm{max}_i$
  ($i=x,y,z$) obtained from the Riemann solver},
  and the {\it transverse} transport velocities at the left
  and right states, for example $\mathcal{V}^y_L$, $\mathcal{V}^z_L$,
  $\mathcal{V}^y_R$, and $\mathcal{V}^z_R$ for the $x$-interface.

\item Obtain a new transverse transport velocity weighting those of the
  left and right states by the \ch{characteristic speeds}
  \ch{\citep{DelZanna2007}}. For the $x$-interface,
  this is
  \begin{equation}
\bar{\mathcal{V}}^{y,z} = \frac{c^\mathrm{max}_x \mathcal{V}_L^{y,z} + 
                                c^\mathrm{min}_x \mathcal{V}_R^{y,z}}
                                {c^\mathrm{max}_x + c^\mathrm{min}_x}\,.
\end{equation}

\item Reconstruct the magnetic fields and the transport velocities to the
  edge where $\mathcal{E}_z$ needs to be calculated. This gives
  $B^{x,y}_{L,R}$ and $\bar{\mathcal{V}}^{x,y}_{R,L}$.

\item Take the maximum \ch{characteristic speeds} in the same way as in step (5) of
  UCT1.

\item Approximate the electric field using the formula by
  \citep{DelZanna2007}:
  \begin{align}
  \label{eq:UCT2}
    \begin{split}
      E_z = \sqrt{\gamma} \left[
        -\frac{c^+_x \bar{\mathcal{V}}^x_L B^y_L
          +c^-_x \bar{\mathcal{V}}^x_R B^y_R
          -c^-_x c^+_x (B^y_R - B^y_L) }
        {c^+_x + c^-_x} \right. \\
        \left.
        +\frac{c^+_y \bar{\mathcal{V}}^y_L B^x_L
          +c^-_x \bar{\mathcal{V}}^y_R B^x_R
          -c^-_y c^+_y (B^x_R - B^x_L) }
        {c^+_y + c^-_y} \right] \,.
    \end{split}
  \end{align}
  Again, we approximate $\mathcal{E}_z=E_z \Delta z$ and use it to
  calculate the circulation.
\end{enumerate}
Since UCT2 appears to be more symmetric than UCT1, we prefer it for the
simulations presented in this work. However, it is worth to mention that
we did not observe important differences, especially no directional bias,
when using with UCT1.

An important difference in our implementation with respect to that of
\citet{londrillo2004divergence} and \citet{DelZanna2007} is that in those
works the quantities $\sqrt{\gamma} B^i$ and not the magnetic fields
$B^i$ are reconstructed at the edges. We instead, factor the square root
of the metric determinant and multiply the whole expression for $E_z$ by
it at the end. This makes the scheme more consistent when evolving a
radial magnetic field in the polar regions, and simplifies the AMR
operations at the poles, as will be seen in section \ref{sec:poles}.

\section{CT-adapted AMR}
\label{sec:AMR-CT}

Constrained transport schemes ensure that, during evolution, no
divergence of the magnetic field is created from one iteration to the
next, at machine precision. However, by construction, no mechanism to
eliminate divergence that has already been created is present.
Therefore, in order to exploit the advantages of constrained transport in
an AMR code, it is necessary to resort to prolongation and restriction
operators that also preserve the constraint $\nabla \cdot \boldsymbol{B}
= 0$ to machine precision. In addition, care must be taken to
synchronise different representations of the same electric and magnetic
field components across fine-coarse interfaces, which is done in a step
similar to the refluxing of finite-volume methods. After defining the
notation in section \ref{sec:notation}, we here describe in detail such
operations.

\subsection{Notation}
\label{sec:notation}

\begin{figure}
\begin{center}
\includegraphics[width=\linewidth]{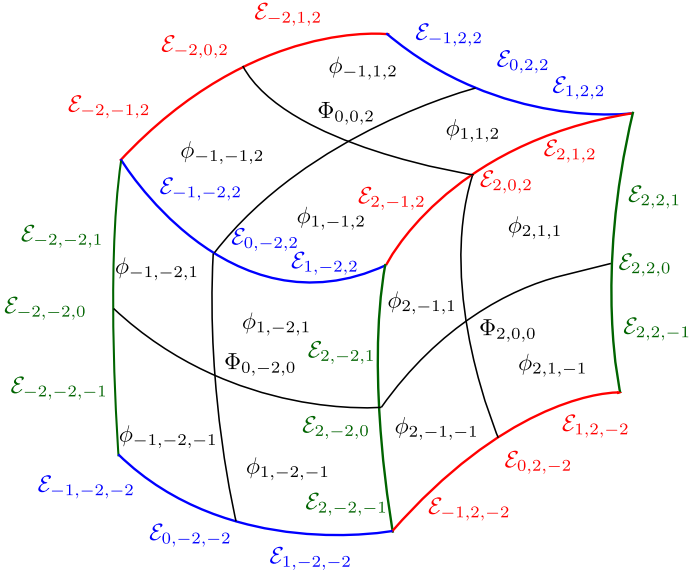}
\end{center}
\caption{Notation for magnetic fluxes and line integrals of the electric
  field for section \ref{sec:notation}.}
\label{fig:amrNotation}
\end{figure}

To avoid using lengthy subscripts, we employ a notation similar to that
used by \citet{Toth2002}. We refer to Figure \ref{fig:amrNotation} for a
schematic view.

The coarse coordinate increment in the direction $i$ is denoted $\Delta
X^i$ and the correspondent fine increment $\Delta x^i=\Delta X^i/2$. The
center of the coarse cell is defined at $(x_0^{1},x_0^{2},x_0^{3})$ and
those of the eight fine cells that result after refinement at $(x_0^{1}
\pm \Delta x^1/2,x_0^{2} \pm \Delta x^2/2,x_0^{3} \pm \Delta
x^3/2)$. Quantities defined at the fine cell centres are labelled by the
subscripts $\pm 1,\pm 1,\pm 1,$ and those at the coarse cell center by
$0,0,0$, although this label is often omitted to keep the notation
simple. Accordingly, $\Delta V$ is the volume of the coarse cell.

The coarse faces defined by the coordinate surfaces of $x^1$ are centered
at $(x_0^{1} + \Delta X^1/2,x_0^{2},x_0^{3})$. Quantities defined
at these faces are labelled by the subscripts $\pm2,0,0$. For instance,
both these faces and their areas are denoted $\Delta S_{\pm2,0,0}$. The
four fine faces in which $\Delta S_{2,0,0}$ is subdivided are labelled
$\Delta s_{2,\pm 1,\pm 1}$. The same applies to $\Delta S_{-2,0,0}$ and
in the other directions. The twelve fine faces $\Delta s_{0,\pm 1,\pm
  1}$, $\Delta s_{\pm 1, 0,\pm 1}$ and $\Delta s_{\pm 1,\pm 1, 0}$ are
not part of any coarse face. The magnetic flux across $\Delta s_{2,1,1}$
is
\begin{equation}
\label{eq:FineFlux}
\phi_{2,1,1} = \int_{\Delta s_{2,1,1}} B^1 \sqrt{\gamma} dx^2dx^3 \,,
\end{equation}
and that across $\Delta S_{2,0,0}$ is
\begin{equation}
\label{eq:CoarseFlux}
\Phi_{2,0,0} = \int_{\Delta S_{2,0,0}}  B^1 \sqrt{\gamma} dx^2dx^3 \,.
\end{equation}
The same applies to quantities defined at cell edges:
$\mathcal{E}_{2,2,0}$ denotes line integral of the electric field along
the coarse edge centered at $(x^1_0+\Delta X^1/2,x^2_0+\Delta
X^2/2,x^3_0)$ and $\mathcal{E}_{2,2,1}$ that along the fine edge centered
at $(x^1_0+\Delta X^1/2,x^2_0+\Delta X^2/2,x^3_0+\Delta x^3/2)$.

In a 2D problem, where symmetry in the $x^3$ direction is assumed, the
integrals with respect to $x^3$ can be ignored, since they only result in
multiplying all quantities by a constant factor that is cancelled in
every equation.

The divergence of the magnetic field in the coarse cell is discretized
as
\begin{equation}
\label{eq:divB}
(\nabla\cdot\boldsymbol{B})=\frac{1}{\Delta V}
\left(\Phi_{2,0,0}-\Phi_{-2,0,0}+\Phi_{0,2,0}-\Phi_{0,-2,0}+\Phi_{0,0,2}-\Phi_{0,0,-2}\right)\,,
\end{equation}
which reduces to the standard definition of divergence
(equation \ref{eq:divB-definition})
in the limit $\Delta V\rightarrow0$, and corresponding
definitions are valid for each of the fine cells.

\subsection{Prolongation and restriction operators for face-allocated variables}
\label{sec:stg-amr}

\subsubsection{Restriction formulas}

As in \citet{Toth2002}, to obtain a restriction formula that naturally
preserves the divergence, we simply require that the flux through one of
the coarse faces is the sum of the fluxes through the fine faces which
form part of it, \ie
\begin{align}
\label{eq:Restriction}
\begin{split}
\Phi_{\pm 2\,0\,0} =& \sum_{j,k=-1,1} \phi_{\pm 2\,j\,k} \,, \\
\Phi_{0\,\pm 2\,0} =& \sum_{i,k=-1,1} \phi_{i\,\pm 2\,k} \,, \\
\Phi_{0\,0\,\pm 2} =& \sum_{i,j=-1,1} \phi_{i\,j\,\pm 2} \,.
\end{split}
\end{align}
The divergence in the newly created coarse cell will be
\begin{equation}
\label{eq:divB_restrict}
(\nabla\cdot\boldsymbol{B})=\frac{1}{\Delta V}
\sum_{i,j,k=-1,1} \Delta V_{i,j,k} (\nabla\cdot\boldsymbol{B})_{i,j,k} \,,
\end{equation}
which is consistent with definition (\ref{eq:divB}) since the interior
fluxes cancel in pairs. If all $(\nabla\cdot\boldsymbol{B})_{i,j,k}$ are
numerical zeros, equation (\ref{eq:divB_restrict}) implies that
$(\nabla\cdot\boldsymbol{B})$ is as well.

\subsubsection{Prolongation formulas}

While divergence-free restriction operators can be derived
straightforwardly as seen above, it is not trivial to find prolongation
operators with the same property. To the best of our knowledge, the only
such prolongation operators for vector fields collocated at cell faces
are those found by \citet{Balsara2001b} and \citet{Toth2002}, who also
derived similar operators for continuous interpolation and for
prolongation of vertex-collocated fields. Since the divergence constraint
alone results in an under-determined system of equations for the fine
magnetic fields, \citet{Toth2002} close the system by requiring also
that, on the fine grid, two different discretisations of the curl in
Cartesian coordinates give the same numerical value. The formulas by
\citet{Balsara2001b} are identical to those of divergence-preserving
continuous interpolation from \citet{Toth2002}, after making consistent
the conventions used in both works for the cell dimensions. However, it
is worth mentioning that \citet{Balsara2001b} obtained them from
different requirements, namely by asking the interpolation of the
magnetic field to be second-order, total-variation-diminishing (TVD), in
addition to have vanishing divergence. The formulas by \citet{Toth2002}
for face-collocated fields and those for continuous interpolation differ
in that the former include additional terms arising from the requirement
of constant discretized curl, while for the latter the curl is allowed to
vary linearly inside the parent cell. \citet{Balsara2004b} provided a
generalised version of his expressions for orthonormal cylindrical and
spherical coordinates, by means of appropriate substitutions in the
Cartesian formula, \eg of the form $B^x \rightarrow rB^r$, $B^y
\rightarrow B^\phi/r$, $B^z \rightarrow B^z$ for cylindrical coordinates.
The aforementioned formulas were successfully implemented and tested for
non-relativistic MHD in the codes \texttt{AstroBEAR}
\citep{Cunninghametal09} and \texttt{BATSRUS} \citep{Toth2012}.

Due to the relation between the curl of the magnetic field and the
electric current, asking the prolonged fields to preserve the value of
the curl is well motivated from a physical point of view. However, this
becomes more complicated when moving to a curved spacetime. On a general
3-dimensional hypersurface, the expressions for the divergence
($\nabla\cdot\boldsymbol{B} = \gamma^{-1/2} \partial_i \gamma^{1/2} B^i$)
and the curl ($(\nabla\times\boldsymbol{B})^i = \epsilon^{ijk}\partial_j
B_k$) of a 3-vector not only involve additional quantities from the
metric, but also operations of raising and lowering indices, or
derivatives with respect to the covariant coordinates. For
non-orthonormal bases, this requires to compute additional approximations
to derivatives or interpolations of the magnetic-field components to
faces in which they were not originally defined, thus increasing the
computational cost and potentially decreasing the accuracy of the
algorithm.

For this reason, we revisit the problem of finding divergence preserving
prolongation operators for face-allocated quantities using an alternative
approach, and derive a general class of such formulas which reduces as a
special case to the operator of \citet{Toth2002} for face-collocated
fields.

Prolongation consists in two steps. First, interpolation is done for the
fine fluxes which are part of a flux on the coarse faces (the {\it
  exterior} fluxes), and then the remaining ({\it interior}) fine fluxes
are computed. Interpolation is done on the magnetic fluxes, not on the
fields, and any formula that satisfies eqs. (\ref{eq:Restriction}) can be
used. The expressions employed in this work are
\begin{align}
\label{eq:exterior_faces}
\begin{split}
\phi_{\pm 2\,j\,k} = \omega\left( \Phi_{\pm 2\, 0 \, 0} + j\frac{\Delta x^2}{2} (\partial_2 \Phi)_{\pm 2 \, 0 \, 0} + k\frac{\Delta x^3}{2} (\partial_3 \Phi)_{\pm 2 \, 0 \, 0}\right)\\
\phi_{i\,\pm 2\,k} = \omega\left( \Phi_{0 \, \pm 2 \, 0} + i\frac{\Delta x^1}{2} (\partial_1 \Phi)_{0 \, \pm 2 \, 0} + k\frac{\Delta x^3}{2} (\partial_3 \Phi)_{0 \, \pm 2 \, 0}\right)\\
\phi_{i\,j\,\pm 2} = \omega\left( \Phi_{0 \, 0 \, \pm 2} + i\frac{\Delta x^1}{2} (\partial_1 \Phi)_{0 \, 0 \, \pm 2} + j\frac{\Delta x^2}{2} (\partial_2 \Phi)_{0 \, 0 \, \pm 2}\right)\,,
\end{split}
\end{align}
where $\omega=1/4$ in 3D and $1/2$ in 2D and $(\partial_i\Phi)$ is any
approximation to the slope of $\Phi$. As \citet{Toth2002}, to compute
these slopes we choose second order formulas such as
\begin{equation}
(\partial_2 \Phi)_{\pm 2\, 0 \, 0} \approx \frac{1}{8 \Delta x^2} \left( \Phi_{\pm 2 \, +4\, 0} - \Phi_{\pm 2 \, -4\, 0} \right) \,.
\end{equation}
However, it is also possible to use limited slopes as \eg {\it minmod},
as suggested by \citet{Balsara2001b}.

Once the fine magnetic fluxes on the exterior faces have been calculated,
the next steps have the purpose of filling the interior fluxes without
creating divergence. We looked for the most general linear formula that
gave these fine interior fluxes in terms of the exterior ones, and which
fulfilled the following requisites:
\begin{enumerate}
\item to keep constant a magnetic flux that was originally constant,
\item to be reversible (specular symmetry), and
\item to preserve the discretized divergence of the magnetic field
in the coarse cell.
\end{enumerate}
First, we write each of interior magnetic fluxes $\phi_{l,m,n}$ as a
combination of the exterior ones and 24 coefficients $n_{i,j,k}$.
\begin{align}
\begin{split}
\phi_{l,m,n}&=\sum_{i=-2,2}\sum_{j,k=-1,1} n_{l,m,n}^{i,j,k} \phi_{i,j,k}
             +\sum_{j=-2,2}\sum_{i,k=-1,1} n_{l,m,n}^{i,j,k} \phi_{i,j,k} \\
            &+\sum_{k=-2,2}\sum_{i,j=-1,1} n_{l,m,n}^{i,j,k} \phi_{i,j,k} \,.
\end{split}
\end{align}
The first requisite can be expressed as
\begin{align}
\label{eq:consistency}
\begin{split}
\phi_{-2,j,k}=\phi_{2,j,k}\,,
\phi_{i,-2,k}=\phi_{i,2,k}\,,
\phi_{i,j,-2}=\phi_{i,j,2} \\
\Rightarrow
\phi_{0,j,k}=\phi_{-2,j,k}\,,
\phi_{i,0,k}=\phi_{i,-2,k}\,,
\phi_{i,j,0}=\phi_{i,j,-2}\,.
\end{split}
\end{align}
Equation (\ref{eq:consistency}) is more general than simply requiring
that a constant magnetic field is kept constant. It means that if the
magnetic flux does not change after entering and exiting the cell, it
will not change inside. This allows to prolong in a consistent way, for
example, a monopolar field in spherical symmetry, for which the magnetic
flux is not the same for all the cell faces at $r=r_0$, but for the
same cell must not
change between the face at $r=r_0$ and that at $r=r_0+\Delta r$. Applied
to the fluxes in the first direction, requirement
eq. (\ref{eq:consistency}) results in 12 equations for the coefficients
of each flux $\phi_{lmn}$, namely
\begin{align}
\label{eq:consistency_n}
\begin{split}
n_{0,m,n}^{-2,j,k}+n_{0,m,n}^{2,j,k}&=\delta^j_m\delta^k_n \,,\\
n_{0,m,n}^{i,-2,k}+n_{0,m,n}^{i,2,k}&=0 \,, \\
n_{0,m,n}^{i,j,-2}+n_{0,m,n}^{i,j,2}&=0 \,,
\end{split}
\end{align}
\noindent
where $\delta^i_j$ is the Kronecker delta.
The next requirement, `reversibility', is the symmetry condition
that when all the exterior fluxes are reflected with respect
to the surface containing $n_{lmn}$, the sign of $n_{lmn}$ must reverse.
This results in another set of relations for the coefficients,
\begin{align}
\label{eq:reversibility}
\begin{split}
n_{0,m,n}^{-2,j,k}&=n_{0,m,n}^{2,j,k}\,,\\
n_{0,m,n}^{-1,\pm 2,k}&=-n_{0,m,n}^{1,\pm 2,k} \,,\\
n_{0,m,n}^{-1,j,\pm 2}&=-n_{0,m,n}^{1,j,\pm 2} \,,
\end{split}
\end{align}
and analogous expression for the other interior fluxes.
Combining with the first requirement (equations \ref{eq:consistency_n}),
and removing the redundant relations, we can already find that
\begin{align}
\label{eq:consistency_and_symmetry}
\begin{split}
n_{0,m,n}^{-2,j,k}&=n_{0,m,n}^{2,j,k}=\frac{1}{2}\delta^j_m\delta^k_n \,,
\\
n_{0,m,n}^{1,2,k} &= - n_{0,m,n}^{1,-2,k} = - n_{0,m,n}^{-1,2,k} =
n_{0,m,n}^{-1,-2,k} \,, \\
n_{0,m,n}^{1,j,2} &= - n_{0,m,n}^{1,j,-2} = - n_{0,m,n}^{-1,j,2} = n_{0,m,n}^{-1,j,-2} \,.
\end{split}
\end{align}
Therefore, only four free parameters remain for determining all the
coefficients for $\phi_{0,m,n}$. These could be, for example
$n_{0,m,n}^{1,2,\pm 1}$ and $n_{0,m,n}^{1,\pm 1,2}$. The same reasoning
leads to similar relations for the coefficients $n_{l,0,n}^{i,j,k}$ and
$n_{l,m,0}^{i,j,k}$, which can be obtained for the other internal fluxes
after cyclic permutations.

Finally, to fulfil the third requirement we ask that the monopolar
magnetic charge (the integral of the divergence) is split evenly between
the eight fine cells. This can be expressed as
\begin{align}
\begin{split}
  (\pm \phi_{\pm 2, j ,k} \mp \phi_{0, j ,k})
&+ (\pm \phi_{i, \pm 2 ,k} \mp \phi_{i, 0, k}) \\
&+ (\pm \phi_{i, j, \pm 2} \mp \phi_{i, j, 0}) = 
\frac{1}{8} \Delta V (\nabla\cdot\boldsymbol{B})\,.
\end{split}
\label{eq:divb_small}
\end{align}
where $\nabla\cdot\boldsymbol{B}$ follows from the discretisation of
equation (\ref{eq:divB}). After substituting the expressions for the
coarse fluxes in terms of the fine ones (equation \ref{eq:Restriction})
and grouping the coefficients associated with each of the exterior fine
fluxes, it is possible to use the independence of these fluxes to obtain
nine equations for the twelve still unknown coefficients in each of the
eight fine cells (more precisely, 24 equations are obtained for each fine
cell, but 12 of them are redundant due to equations
(\ref{eq:consistency_and_symmetry}) and three are not independent from
the others, leaving nine equations). Therefore, this constraint leaves
three free parameters for each of the eight fine cells.

\begin{figure}
\centering \includegraphics[width=\linewidth]{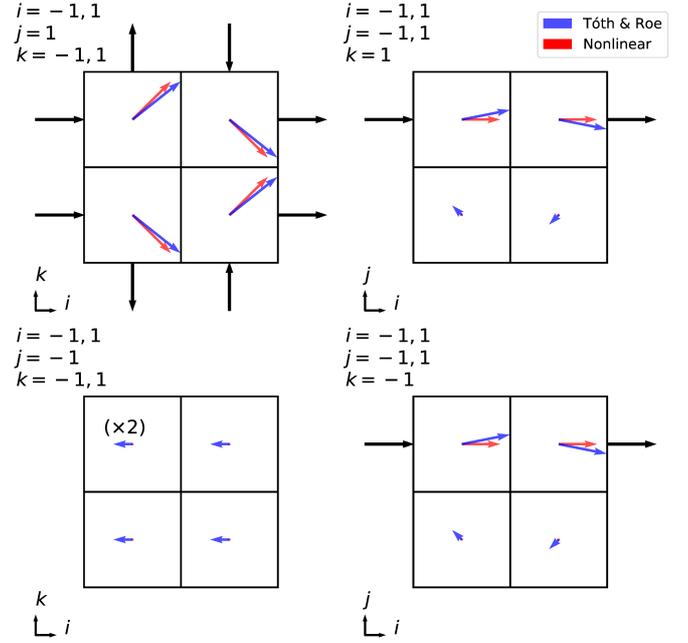}
\caption{Comparison between the \ch{different projections of the}
  interior fields obtained using \ch{T\'oth \& Roe's operator (blue) and
    the nonlinear operator here derived (red), from the same exterior
    fluxes (black) and for a cubic cell.  Each panel corresponds to a
    slice of a coarse cell, for which the indices of the fine cells are
    given at the upper left (\cf Figure \ref{fig:amrNotation}). The
    direction of the original flux is better preserved by the nonlinear
    operator (\cf right panels), where a component in the $y$-direction
    that was not originally present appears for the T\'oth \& Roe
    operator.  The field interpolated using the nonlinear operator is
    exactly zero for cells with $j=-1$.} For clarity, vectors in the
  lower left panel are represented with a double size.}
\label{fig:test_cubes}
\end{figure}

Since each of the internal faces is shared by two fine cells and its free
parameters need to be compatible on both sides, it turns out that all of
the coefficients of the internal fluxes can be written in terms of the
same three parameters, which we name $\alpha_1$, $\alpha_2$ and
$\alpha_3$. For example, the interior magnetic fluxes in direction $x^1$
can be obtained in a divergence-free way as
\begin{align}
\label{eq:interior_faces}
\begin{alignedat}{2}
\phi_{0\,-1\,-1} =& \frac{1}{2}\left(\phi_{-2\,-1\,-1} + \phi_{2\,-1\,-1}\right) \\
                 +& \frac{1}{16} [(3+\alpha_2) F_{1\,2\,-1} + (1-\alpha_2) F_{1\,2\,1} \\
                  & \quad + (3-\alpha_3) F_{1\,-1\,2} + (1 + \alpha_3) F_{1\,1\,2} ]\\
\phi_{0\,-1\,1} =& \frac{1}{2}\left(\phi_{-2\,-1\,1} + \phi_{2\,-1\,1}\right)\\
                +& \frac{1}{16} [(1-\alpha_2) F_{1\,2\,-1} + (3+\alpha_2) F_{1\,2\,1} \\
                 & \quad + (3-\alpha_3) F_{1\,-1\,2} + (1 + \alpha_3) F_{1\,1\,2} ]\\
\phi_{0\,1\,-1} =& \frac{1}{2}\left(\phi_{-2\,1\,-1} + \phi_{2\,1\,-1}\right)\\
                +& \frac{1}{16} [(3+\alpha_2) F_{1\,2\,-1} + (1-\alpha_2) F_{1\,2\,1} \\
                 & \quad + (1+\alpha_3) F_{1\,-1\,2} + (3 - \alpha_3) F_{1\,1\,2} ]\\
\phi_{0\,1\,1} =& \frac{1}{2}\left(\phi_{-2\,1\,1} + \phi_{2\,1\,1}\right)\\
               +& \frac{1}{16} [(1-\alpha_2) F_{1\,2\,-1} + (3+\alpha_2) F_{1\,2\,1} \\ 
                & \quad + (1+\alpha_3) F_{1\,-1\,2} + (3 - \alpha_3) F_{1\,1\,2} ] \,,
\end{alignedat}
\end{align}
where 
\begin{align}
\label{eq:interior_F}
\begin{alignedat}{7}
&F_{1\,2\,-1}&=&\phi_{1\,2\,-1}&-&\phi_{1\,-2\,-1} &-&\phi_{1\,2\,-1}& +&\phi_{-1\,-2\,-1}&\\
& F_{1\,2\,1}&=& \phi_{1\,2\,1}&-&\phi_{1\,-2\,1}  &-&\phi_{-1\,2\,1}& +&\phi_{-1\,-2\,1}&\\
&F_{1\,-1\,2}&=&\phi_{1\,-1\,2}&-&\phi_{1\,-1\,-2} &-&\phi_{-1\,-1\,2}&+&\phi_{-1\,-1\,-2}&\\
& F_{1\,1\,2}&=& \phi_{1\,1\,2}&-&\phi_{1\,1\,-2}  &-&\phi_{-1\,1\,2}& +&\phi_{-1\,1\,-2}&\,.
\end{alignedat}
\end{align}
Corresponding formulas for the other directions can be obtained by
circular permutation of direction of the fluxes and of the indices of
$\alpha_i$s. Since the parameters $\alpha_i$ affect only the interior
fluxes, they may vary from cell to cell depending \eg on the cell
geometry. For example, the formulas reduce to those by \citet{Toth2002}
when
\begin{equation}
\alpha_1 = \frac{\Delta y^2 - \Delta z^2}{\Delta y^2 + \Delta z^2} \,,
\alpha_2 = \frac{\Delta z^2 - \Delta x^2}{\Delta z^2 + \Delta x^2} \,,
\alpha_3 = \frac{\Delta x^2 - \Delta y^2}{\Delta x^2 + \Delta y^2} \,,
\end{equation}
\noindent
\ch{where} we rename $\Delta x^1\rightarrow x$, $\Delta x^2 \rightarrow y$
and $\Delta x^3 \rightarrow z$ to keep the notation simple.
This choice has a directional bias which depends on the geometry of
the cell, and which disappears when $\Delta x = \Delta y = \Delta z$ and
thus $\alpha_i=0$, which gives the form of the operator used by
\citet{Cunninghametal09}. Another feature of this choice, and for any
other in which $\alpha_i$ do not depend on the exterior magnetic fluxes,
is that, prolongation can produce a $z$ component in a field originally
having only $x$ and $y$ components.
We noticed that this can be avoided at the expense of making the
operator nonlinear, by resorting to a heuristic formula
for $\alpha_i$ to control the magnetic flux mixing. First, we quantify
this mixing in the $z$ direction as
\begin{equation}
\sigma_z=\frac{\sum_{\mathrm{up}}|\phi| - \sum_{\mathrm{down}}|\phi|}{\sum_{\mathrm{up}}|\phi| + \sum_{\mathrm{down}}|\phi|} \,,
\end{equation}
where $\sum_{\mathrm{up}}$ runs on the exterior fluxes in the $x$ and $y$
direction residing in the upper half of the coarse cell, and
$\sum_{\mathrm{down}}$ on those residing in the lower half.
Corresponding expressions are defined for quantifying the mixing in the
$x$ and $y$ directions.
Then, we set the $\alpha_i$-functions for interpolation as
\begin{align}
\alpha_1=\sigma_y - \sigma_z \,,
\alpha_2=\sigma_z - \sigma_x \,,
\alpha_3=\sigma_x - \sigma_y \,.
\end{align}
By substituting, it is possible to verify that any magnetic-field
configuration that originally had no components in a given coordinate
direction will not show them when prolonged. A comparison between the
interior fields interpolated using this nonlinear operator and that by
T\'oth \& Roe is shown in Figure \ref{fig:test_cubes}.

\subsection{Restriction of edge-allocated variables}
\label{sec:edge_restriction}

On each block, magnetic fluxes are updated using the representation of
$\mathcal{E}$ correspondent to the level of the block. However, at a
fine/coarse interface two different representations of the magnetic flux
exist. On the coarse side we have, for example, the magnetic flux
$\Phi_{2,0,0}$, while on the fine side, we have
$\sum_{i,j=-1,1}\phi_{2,j,k}$. Due to the use of different
representations of $\mathcal{E}$ for the update, the two representations
of the magnetic flux differ, although in absence of surface monopolar
magnetic charge they should coincide. In order to make them consistent
without creating divergence on the coarse side, we replace the coarse
representation of $\mathcal{E}$ used in the calculation of the
circulation with the fine one. To do this, we store both representations
of $\mathcal{E}$ on the interface at the time they are calculated. After
the time update, we subtract the line integrals of the electric field
contained in the interface from the fluxes in contact with it. We then
communicate the fine representations to the coarse side and construct
a new coarse representation. For example, for an interface at
constant $x^1$ between a coarse region at the left and a fine region
at the right, we replace
\begin{align*}
\mathcal{E}_{2,2,-1} + \mathcal{E}_{2,2,1} &\rightarrow \mathcal{E}_{2,2,0}\\
\mathcal{E}_{2,-1,2} + \mathcal{E}_{2,1,2} &\rightarrow \mathcal{E}_{2,0,2}\\
\mathcal{E}_{2,-2,-1} + \mathcal{E}_{2,-2,1} &\rightarrow \mathcal{E}_{2,-2,0}\\
\mathcal{E}_{2,-1,-2} + \mathcal{E}_{2,1,-2} &\rightarrow \mathcal{E}_{2,0,-2}
\end{align*}
for each coarse cell and recalculate the circulation to update all the fluxes
except for $\Phi_{-2,0,0}$. In 2D, no sum is necessary and the coarse
representations are simply replaced by the co-spatial fine ones.

%
%

\subsection{Parallelisation and ghost-cell exchange}
\label{sec:gc-exchange}

\begin{figure*}
\begin{center}(a)
\includegraphics[height=4cm]{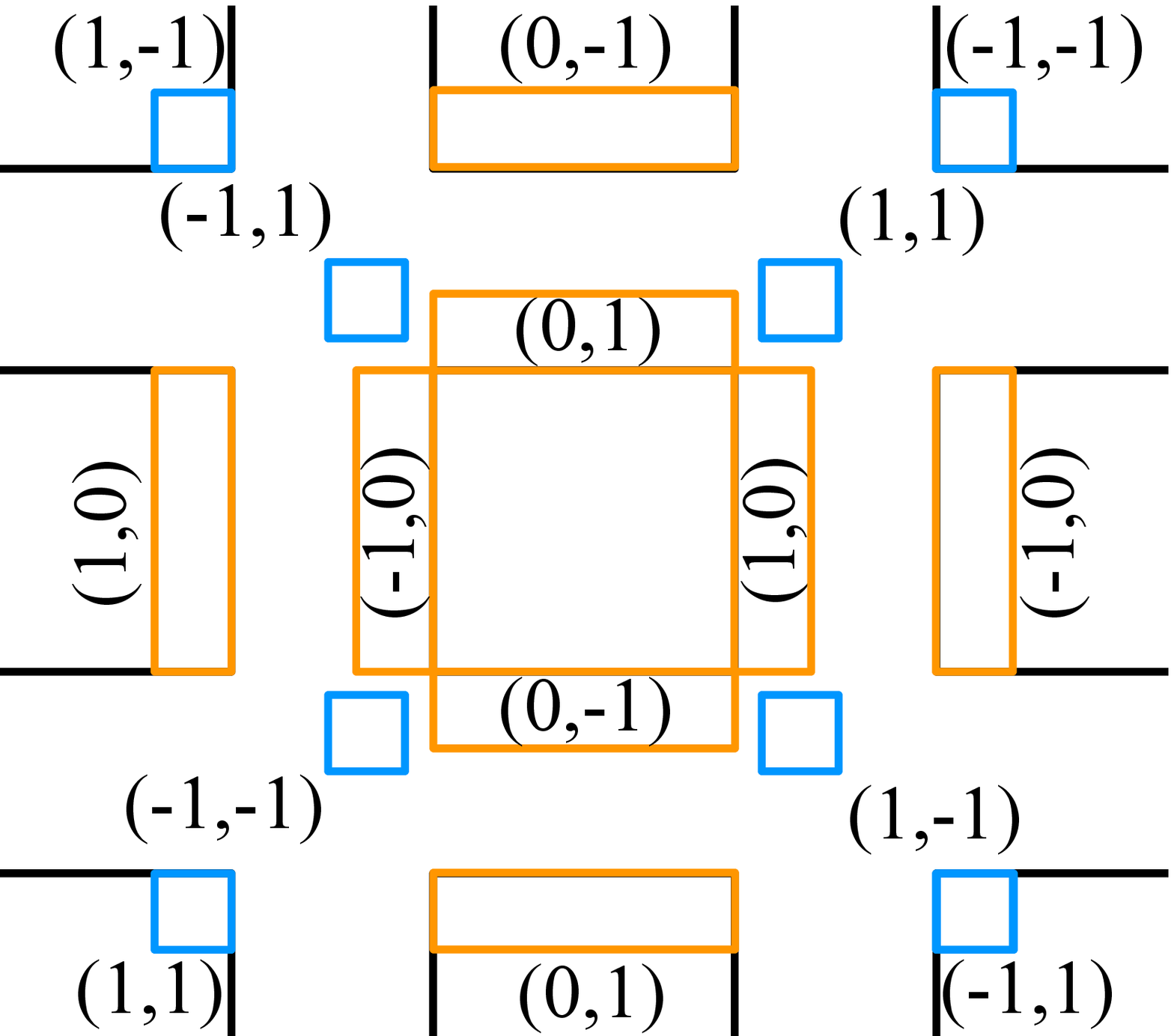}
\hspace{0.5cm}(b)
\includegraphics[height=4cm]{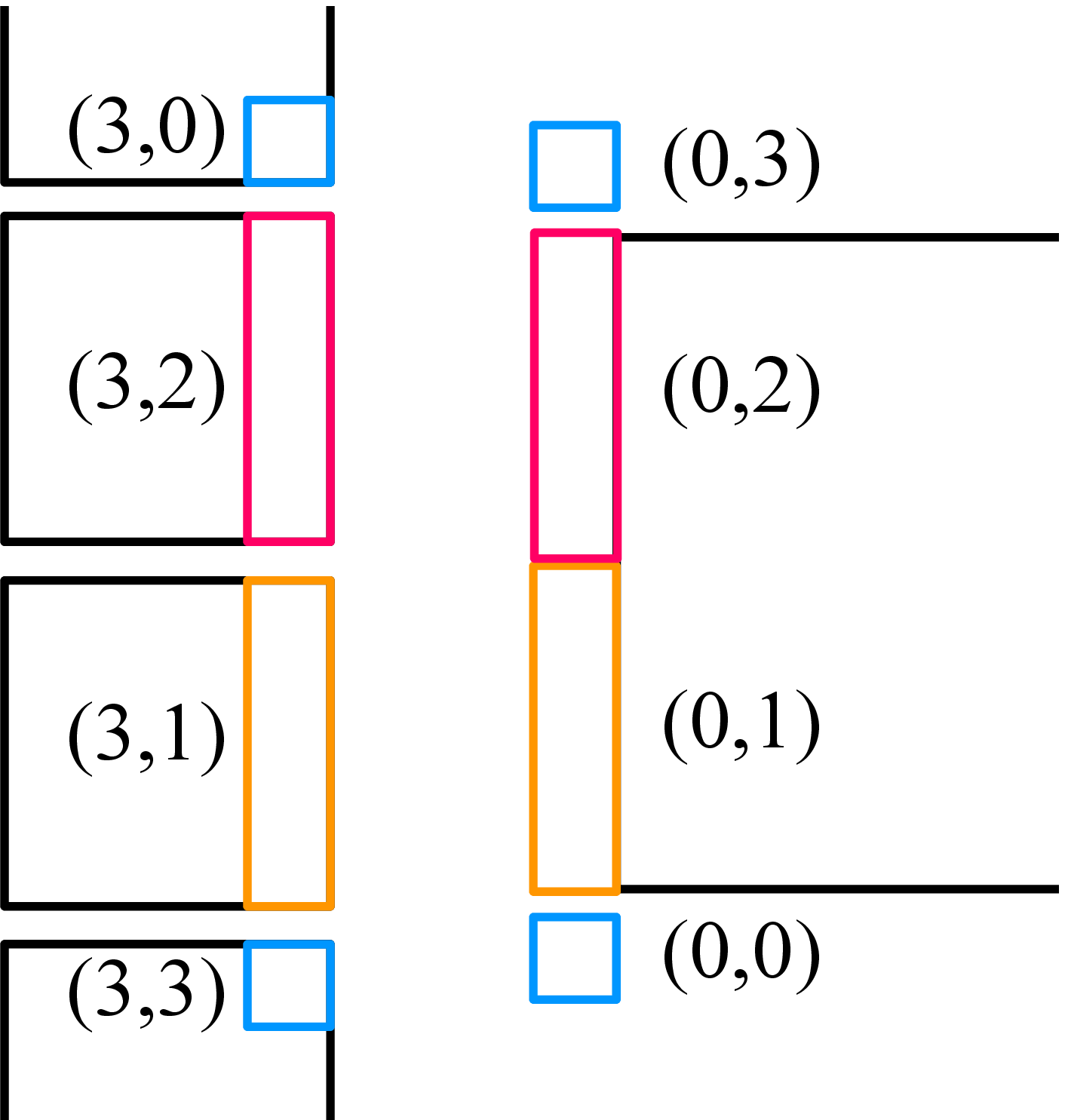}
\hspace{0.5cm}(c)
\includegraphics[height=5cm]{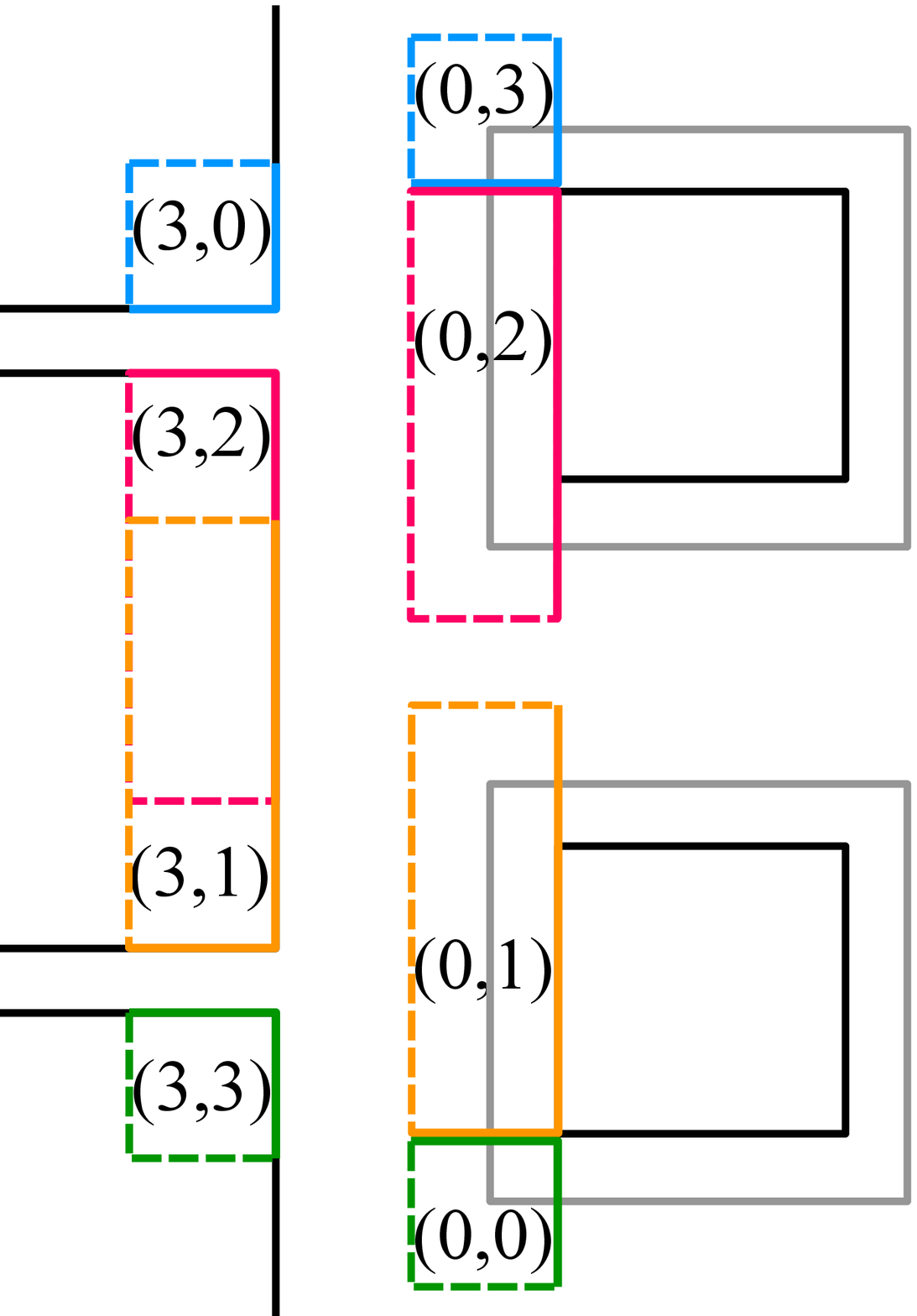}
\end{center}
\caption{Ghost-cell exchange for blocks (a) at the same resolution level
  and for coarse/fine interfaces in (b) restriction and (c) prolongation.
  The numbers represented are the local grid indices to identify
  neighbours and ghost regions, as explained in Section
  \ref{sec:gc-exchange}. Lines drawn in colour denote (magnetic) fluxes
  on the boundary faces that are communicated.
  In panel (c), the dashed lines denote boundaries that
  contain only tangential fluxes necessary for restriction which reside
  only on the {\it coarse} buffer of a block. The size of fine blocks
  including ghost regions is marked by a grey line.}
\label{fig:gc_types}
\end{figure*}

To simplify ghost-cell communications in the code, a block's
neighbours and its children are identified by local grid indices
$\texttt{i1},\texttt{i2},\texttt{i3}=\{-1,0,1\}$ and local child grid
indices $\texttt{ic1},\texttt{ic2},\texttt{ic3}=\{0,1,2,3\}$, both from
left (back, bottom) to right (front, top). For example, if block $A$ is
directly at the left of block $B$ both belong to the same resolution
level, block $A$ is identified as neighbour $(-1,0)$ of block $B$, and
block $B$ is identified as neighbour $(1,0)$ of block $A$. This
identification strategy is inherited from \texttt{MPI-AMRVAC} \citep[see
  \eg Figure 4 of][]{Keppens2012}. When exchanging ghost cells, different
array ranges are associated to each index combination, corresponding to
the extent of the ghost zones that need to be filled from that
neighbour. The identification of neighbours and ghost zones with integer
indices allows to easily determine communication patterns using integer
arithmetic operations. As an example, Figure \ref{fig:gc_types} shows some
of these communication patterns for the case of a two-dimensional grid.

The three panels of Figure \ref{fig:gc_types} show examples of how these
local indices are used to identify different ghost regions for (a) same
resolution level communications, (b) restriction and (c) prolongation.
To ensure that the communicated values are consistent, first same
resolution communications are performed, then restrictions and finally
prolongations. As in \texttt{MPI-AMRVAC}, to minimise the size of
communications at resolution jumps, variables are restricted before being
sent and are prolonged after being received. Each block possess a buffer
containing a coarse representation of it to facilitate this operations.

In contrast to \texttt{MPI-AMRVAC}, for which inter-processor ghost-cell
communications were based on MPI types created using the functions
\texttt{MPI\_CREATE\_SUBARRAY} and \texttt{MPI\_TYPE\_COMMIT}
\citep[see][]{Keppens2012}, \texttt{BHAC} packs all communications, both
for staggered and cell centered variables, in one-dimensional arrays
using FORTRAN's {\tt reshape} function. These buffer arrays are
transmitted to other processors using non-blocking MPI communications
through the functions {\tt MPI\_ISEND} and {\tt MPI\_IRECV}. When
received at the target destination, they are unpacked to fit the shape of
the ghost region that they need to fill.

A reason for this change is that the array segments that need to be
communicated for each component of the staggered field are different
among them and different from those of the cell-centered variables;
therefore, using a single buffer instead of a different {\tt MPI\_TYPE}
for each of them reduces the number of necessary communications. These
and other changes done in the ghost-cell exchange routines are oriented
to a future upgrade of the code to a task-based scheduling, which would
result in a significant speedup and facilitate the use of hierarchical
time-stepping.

Finally, fluxes and electric fields necessary for the consistency steps
described in section \ref{sec:edge_restriction} are also written in
special communication buffers for sending and receiving. These buffers
are likewise one-dimensional arrays filled and extracted using the
Fortran \texttt{reshape} function.

\subsection{Poles in spherical and cylindrical coordinates}
\label{sec:poles}

\begin{figure}[h]
\includegraphics[width=1.0\linewidth]{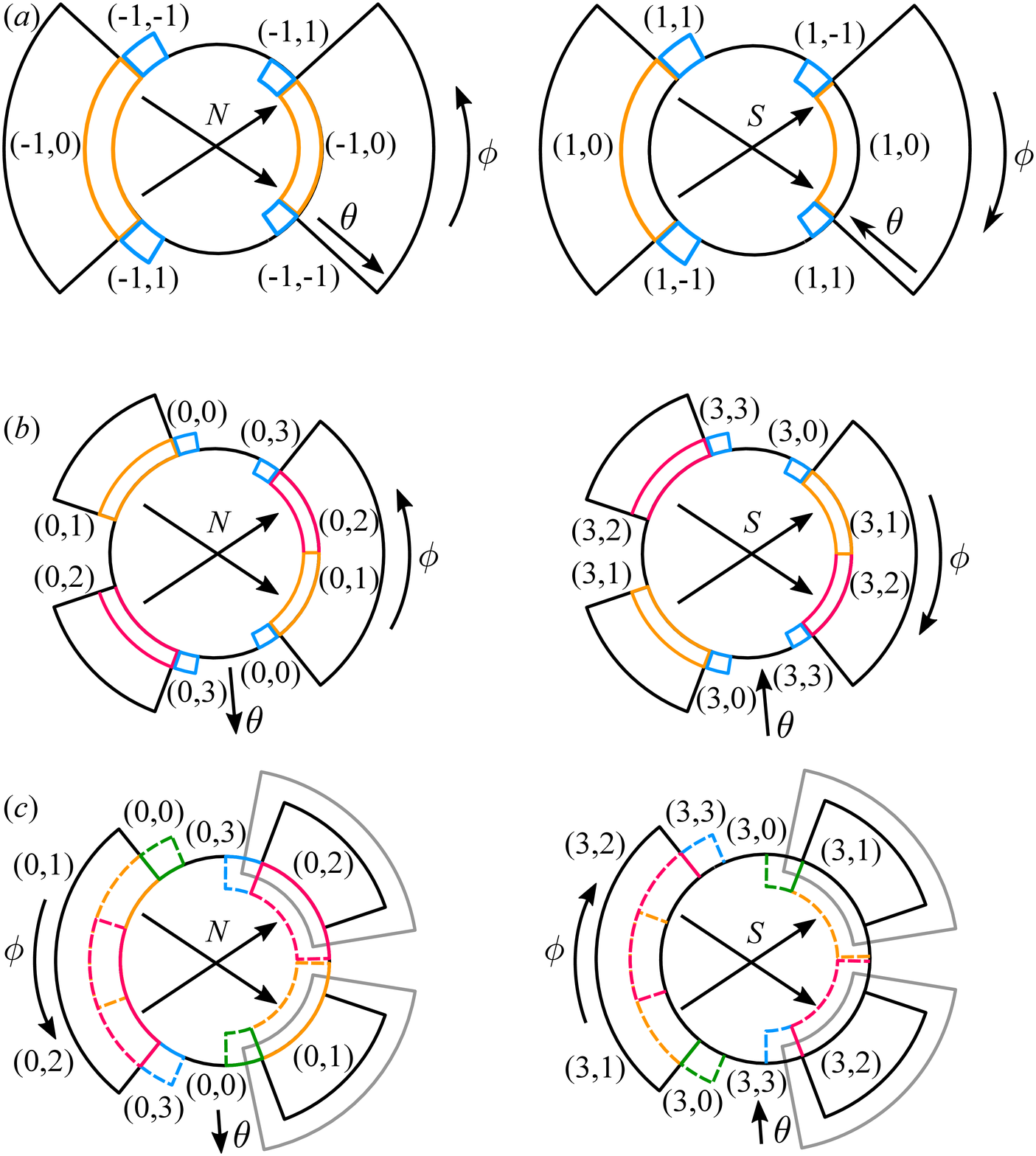}
\caption{\label{fig:gc_poles} Ghost-cell exchange at the south and north
  poles in spherical coordinates for same resolution level (a),
  restriction (b) and prolongation (c). For clarity, the pole is
  represented as an expanded black circumference to identify
  the cell faces touching it, although no there is no excised region.
  As in Figure \ref{fig:gc_types}, colored lines denote (magnetic) fluxes
  on the boundary faces that are communicated.
  For prolongation, dashed lines denote
  boundaries that contain only tangential fluxes necessary for
  restriction and the total extent of fine blocks including ghost regions
  is marked by a grey line. For cylindrical coordinates, the ghost-cell
  exchange at the $z$ pole happens as for the north pole.}
\end{figure}

In three-dimensional simulations in spherical and cylindrical geometry,
the ghost-cell exchange at the poles requires special attention. Despite
being located at the minimum (and maximum) $\theta$ for spherical and
$\rho$ for cylindrical coordinates, in \texttt{BHAC} the pole is not
treated as a physical boundary nor it is excised as is usually done
in many codes \citep[see \eg][]{Shiokawa2012}. Instead, the
connectivity of the blocks touching the poles is changed so that
each on them considers as its neighbours the blocks at the opposite
side of the polar axis and reconstructions can be performed across
the latter.

The only difference respect to the usual ghost-cell exchange is that,
after being received, the elements of the receive buffer are reversed in
the $\phi$ direction, as is shown in Figure \ref{fig:gc_poles}.
Likewise, periodic boundary conditions such as those in the $\phi$
direction, are handled by changing the connectivity of the grid such that
the first block in the $\phi$ direction becomes the neighbour of the last
one.

The flux-fixing operation is not necessary at poles, since the areas of
cell `surfaces' touching the pole are exactly zero and no flux should be
present. Similarly, the contribution of the circulation of the electric
field is exactly zero along the polar axis, by recalling that in a
coordinate basis $\sqrt{\gamma}$ vanishes there
(see equations \ref{eq:UCT1} and \ref{eq:UCT2}). Therefore, no fixings
are performed and both numerical fluxes and electric fields are set
to zero at the pole.

\subsection{Boundary conditions}

When non-periodic boundary conditions are used, it is necessary to ensure
that no divergence is created at boundaries. To accomplish this, we
first extrapolate the components of the magnetic field tangential to the
boundary according to a user-given prescription, \eg symmetric,
antisymmetric, flat, etc. Then, the normal component is filled layer by
layer outwards from the boundary, cancelling the sum of the magnetic
fluxes for each cell. For symmetric and antisymmetric boundary
conditions, this procedure can be applied without modifications also at
resolution jumps; however, this is not the case for flat boundary
conditions. The reason is that they consist in copying the value of
a variable at the last cell of the physical domain to all
of the ghost cells; and in a resolution
jump, the sum of two times the last magnetic flux on the fine side is not
necessarily equal to the last magnetic flux on the coarse side.
Therefore, instead of filling the ghost cells with the values of the last
row in the domain, we fill them with an average of the last {\it two}
rows, weighted by the normal surfaces. In this way, the fine and the
coarse representations of the fields at the ghost cells are consistent.

\section{Numerical tests}
\label{sec:tests}

In this section we describe the results of three well known numerical
tests performed to validate the new schemes, two in special relativity
and one in general relativity. These are the Gardiner \& Stone advected
magnetic loop \citep{Gardiner2005}, the cylindrical explosion of
\citet{Komissarov1999}, and the highly magnetized stationary torus by
\citet{Komissarov2006a}. Since the new additions (CT) and AMR can only
be tested in 2D and 3D, and the code has already been validated for 1D
problems \citep[see][]{Porth2017}, we omit 1D tests. We also omit tests
using \ch{both the cell-centred and the staggered variants of
arithmetic averaging, except for comparisons with UCT, as the former} have
already been published in \ch{\citet{Porth2017} and} \citet{Olivares2018a}.

We are therefore interested in testing the UCT algorithms and the new AMR
features, as well as the recently implemented limiters, WENO Z+ and MP5.
All of the tests presented here use the equation of state of an ideal
fluid with $\hat{\gamma}=4/3$. The prescription for triggering AMR is
the L\"ohner scheme \citep{Loehner87}, which decides to coarsen or
refine based on a quantification of oscillations in the numerical
solution.

\subsection{Loop advection}
\label{sec:loop}

\begin{figure*}
\centering
\includegraphics[width=\linewidth]{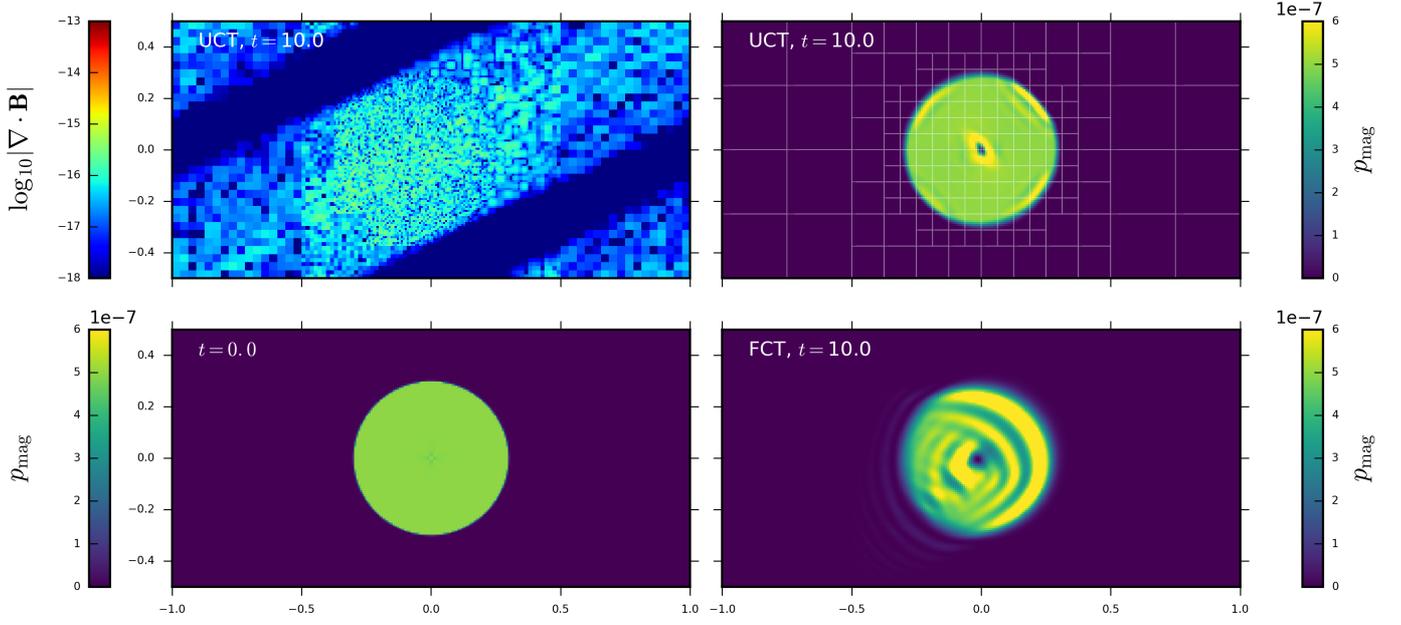}
\caption{Gardiner \& Stone's loop advection test. {\it Top left:}
  Divergence of the magnetic field after ten time units, using UCT2.
  {\it Bottom left:} Magnetic pressure at the initial configuration.
  {\it Top right:} After ten time units, using UCT2, showing the AMR
  blocks of $8\times 8$ cells.
  {\it Bottom right:} After ten time units, using flux-CT. }
\label{fig:loop-test}
\end{figure*}

A well known problem to assess the importance of spurious effects due to
the divergence-control technique employed in an MHD code is the advection
of a weakly magnetized loop. This test was originally used by
\citet{Gardiner2005} to study a divergence-preserving scheme with the
upwind property. In this section, we present a comparison of the results
obtained using the cell-centered version of flux-CT and the upwind method
UCT2.

On a uniform background with $\rho=1$, $p=1$, $v^x=0.2$, and $v^y=0.1$,
the problem consists on advecting a magnetic loop with radius $R=0.3$,
described by the potential $A_z = A_0(R - r)$ for $r\leq R$ and $A_z = 0$
for $r>R$, where $r^2=x^2+y^2$. $A_0$ is chosen $10^{-3}$, so that
$\beta=p_\mathrm{fluid}/p_\mathrm{mag}=10^6$ and the loop is advected
passively. The domain is $x\in [-1,1]$ and $y\in [-0.5,0.5]$ with
periodic boundary conditions. The methods employed are the MP5 limiter,
HLLE fluxes and the RK3 time-integrator.

Two realisations of the same set-up are evolved: one using the flux-CT
method, and the other UCT2. While for the former it is not possible to
use AMR, for the latter three levels of AMR are used. The highest
AMR-level has a resolution equivalent to that of the flux-CT run, i.e,
$256\times128$.

The magnetic pressure at the initial condition and some snapshots of the
evolution are depicted in figure \ref{fig:loop-test}. The upper-left
panel of the same Figure shows the divergence of the magnetic field for
the UCT2 run at the final time $t=10.0$, when the magnetic loop, and thus
the refined region, has travelled a complete cycle across the domain.
Divergence remains always zero at machine precision ($\sim 10^{-16}$)
despite resolution jumps and coarsening/refining of blocks. It can be
seen that, in addition to the AMR capacity, the staggered UCT scheme is
able to preserve much better the original shape of the loop, without
creating as many spurious oscillations as flux-CT.

\subsection{Cylindrical explosion}
\label{sec:cylindrical_explosion}

\begin{figure}
\includegraphics[width=\linewidth]{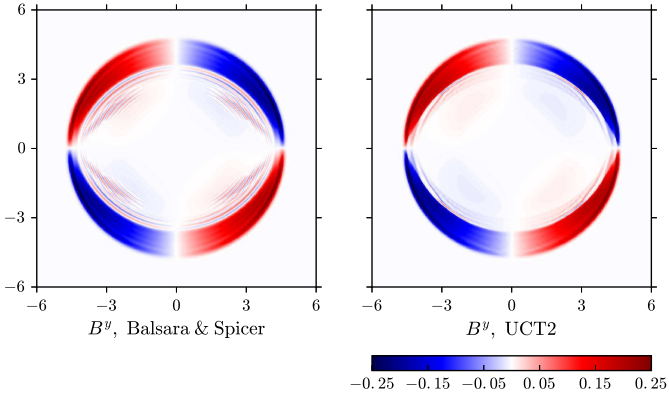}
\caption{Cylindrical explosion. $y$-component of the
magnetic field at $t=4.0$ when using the BS algorithm
({\it left})
and UCT2 ({\it right}) \label{fig:Cyl_explosion_snapshot}}
\end{figure}

\begin{figure}
\includegraphics[width=\linewidth]{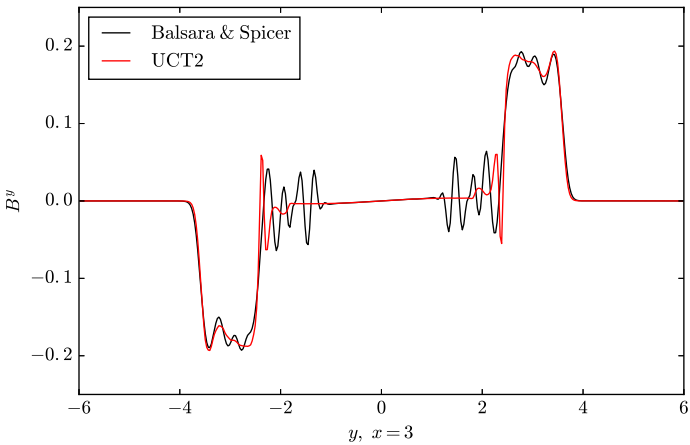}
\caption{Cylindrical explosion. Profile of the $y$-component
  of the magnetic field along a line at
  $x=3$. The Balsara \& Spicer algorithm shows a more oscillatory
  behavior as compared with UCT2. \label{fig:Cyl_explosion_profile}}
\end{figure}

Another challenging problem for MHD codes is that of a cylindrical blast
wave expanding in a homogeneous plasma with an initially constant
magnetic field. We present results of such problem, which highlights the
importance of upwinding in the calculation of electric fields, and show
the different behavior of the \ch{BS} and the
UCT2 schemes. The problem set-up is the same that appears in
\citet{DelZanna2007} and \citet{Komissarov1999}.

The domain extends over $[-6,6]\times[-6,6]$. Plasma is initially at
rest, with a constant magnetic field $B^x=0.1$, $B^y=B^z=0$. Defining
$r^2=x^2+y^2$, the domain is divided in three regions: a high density and
pressure ($\rho_\mathrm{in}=10^{-2}$, $p_\mathrm{in}=1$) region for
$r<r_\mathrm{in}=0.8$, a low density and pressure
($\rho_\mathrm{out}=10^{-4}$, $p_\mathrm{out}=5\times 10^{-4}$) region
for $r>r_\mathrm{out}=1.0$, and a transition region for
$r_\mathrm{in}\leq r \leq r_\mathrm{out}$ where
$\rho=\rho_\mathrm{in}(r/r_\mathrm{in})^{\alpha_1}$ and
$p=p_\mathrm{in}(r/r_\mathrm{in})^{\alpha_2}$, and the exponents
$\alpha_1$ and $\alpha_2$ are such that $p(r)$ and $\rho(r)$ are
continuous at $r_\mathrm{in}$ and $r_\mathrm{out}$. The boundary
conditions are periodic, although this is not relevant in practice, since
the region of interest cannot be affected by signals coming form the
boundaries during the simulation time.

The system is evolved until $t=4.0$ with an RK3 integrator and a CFL
factor of 0.35, and HLLE fluxes with WENOZ+ reconstruction are used. The
simulation used three AMR levels, with a base resolution of
$100\times100$, giving an effective resolution of $400 \times 400$. To
compare the two algorithms for magnetic-field evolution, we run two
simulations, one using the \ch{BS} algorithm and the other using UCT2.

Figure \ref{fig:Cyl_explosion_snapshot} shows the last snapshot of the
evolution for both methods. Spurious oscillations behind the reverse
shock are more pronounced for the \ch{BS} method, similar to
those already encountered in Section \ref{sec:loop}.

The situation greatly improves when adopting the UCT2 algorithm, as can
be further confirmed by examining the profiles of the $y$-component
of the magnetic field along the line $x=3$ shown in
Figure \ref{fig:Cyl_explosion_profile}.

\subsection{Stationary torus}

The test presented in this section is the analytic solution found by
\citet{Komissarov2006a} for an equilibrium torus with a toroidal magnetic
field. As in \citet{Porth2017}, the implementation of the initial
condition in \texttt{BHAC} is based on routines kindly provided by Chris
Fragile. Taking advantage of the stationarity of this analytic solution,
we use this test to quantify the convergence of the numerical solution.
Recalling the results from \citet{Komissarov2006a}, for a fluid of
constant angular momentum distribution $l=-u_\phi/u_t = l_0$, and
adopting a specific relation of the fluid and magnetic pressure with
respect to the fluid enthalpy, the configuration is described by the
equation
\begin{align}
\mathcal{W} - \mathcal{W}_{in} = - {\kappa \over \kappa -1} {p \over \rho h}
 - {\eta \over \eta-1} {p_{m} \over \rho h}  \,,
\end{align}
where $\mathcal{W} = \ln |u_t|$ and $\kappa$ and $\eta$ are constants. To
completely specify the solution, it is necessary to give two more
parameters, namely the fluid enthalpy and the ratio of fluid to magnetic
pressure at the centre of the torus, $(\rho h)_c=\omega_c$ and $(p/p_{\rm
  mag})_c = \beta_c$, respectively. The parameters of the torus chosen
here as a test problem are $\kappa = 4/3$, $\eta = 4/3$, $l_0 = 2.8$,
$r_c = 4.62$, $\mathcal{W}_{in} = -0.030$, $\omega_c = 1.0$, and $\beta_c
= 0.1$. The dimensionless spin of parameter of the black hole is set to
$a=0.9$, and the plasma follows the equation of state of an ideal fluid
with $\hat{\gamma}=4/3$. Simulations are performed in 3D and using
logarithmic Kerr-Schild coordinates (\ie MKS coordinates with $\chm{\vartheta_0}=0$ and
$R_0 =0$). The simulation domain spans over $\theta\in[\pi/5,4\pi/5]$,
$r_{\rm KS}\in[0.95\, r_{\mathrm{h}},\ 50\,M]$ and $\phi \in [0,\pi]$,
where $r_{\rm KS}$ is the radial Kerr-Schild coordinate and
$r_{\mathrm{h}}$ is the (outer) event horizon radius of the BH. To test
convergence, we performed three simulations progressively doubling the
resolution from a base grid of $N_r\times N_\theta \times N_\phi = 100
\times 60 \times 40$. The rotation period of the center of the torus is
$68\,M$, and the simulation is carried out until $t=100\,M$, thus
spanning nearly one and a half orbital periods. Although this kind of
solutions are known to be unstable to the MRI and the Papaloizou-Pringle
instability in 3D \citep{Wielgus2015,Bugli2018}, the time elapsed by our
simulations still corresponds to the initial phase of the linear growth,
and no chaotic behaviour can be observed. In fact, the torus here
presented is identical to Case A simulated by \citet{Wielgus2015}, who
reported perturbations at $t=100\ M$ still at the $\sim10^{-4}$ level.


To emulate vacuum regions outside the torus, a tenuous atmosphere with
density and pressure given by $\rho=\rho_{\mathrm{min}} \, r_{\rm
  KS}^{-3/2}$, $p=p_{\mathrm{min}} \, r_{\rm KS}^{-5/2}$ with
$\rho_{\mathrm{min}}=10^{-5}$ and $p_{\mathrm{min}}=10^{-7}$ was added.
Initially, the velocity of the atmosphere is set to zero in the Eulerian
frame. However, the atmosphere is free to evolve, and only density and
pressure are reset to the atmosphere value whenever they fall below it.
The numerical methods chosen for this test were the HLLE Riemann solver,
LIMO3 reconstruction and two-step time integration with CFL factor $0.5$.

\begin{figure*}
\centering
\includegraphics[width=0.3\linewidth]{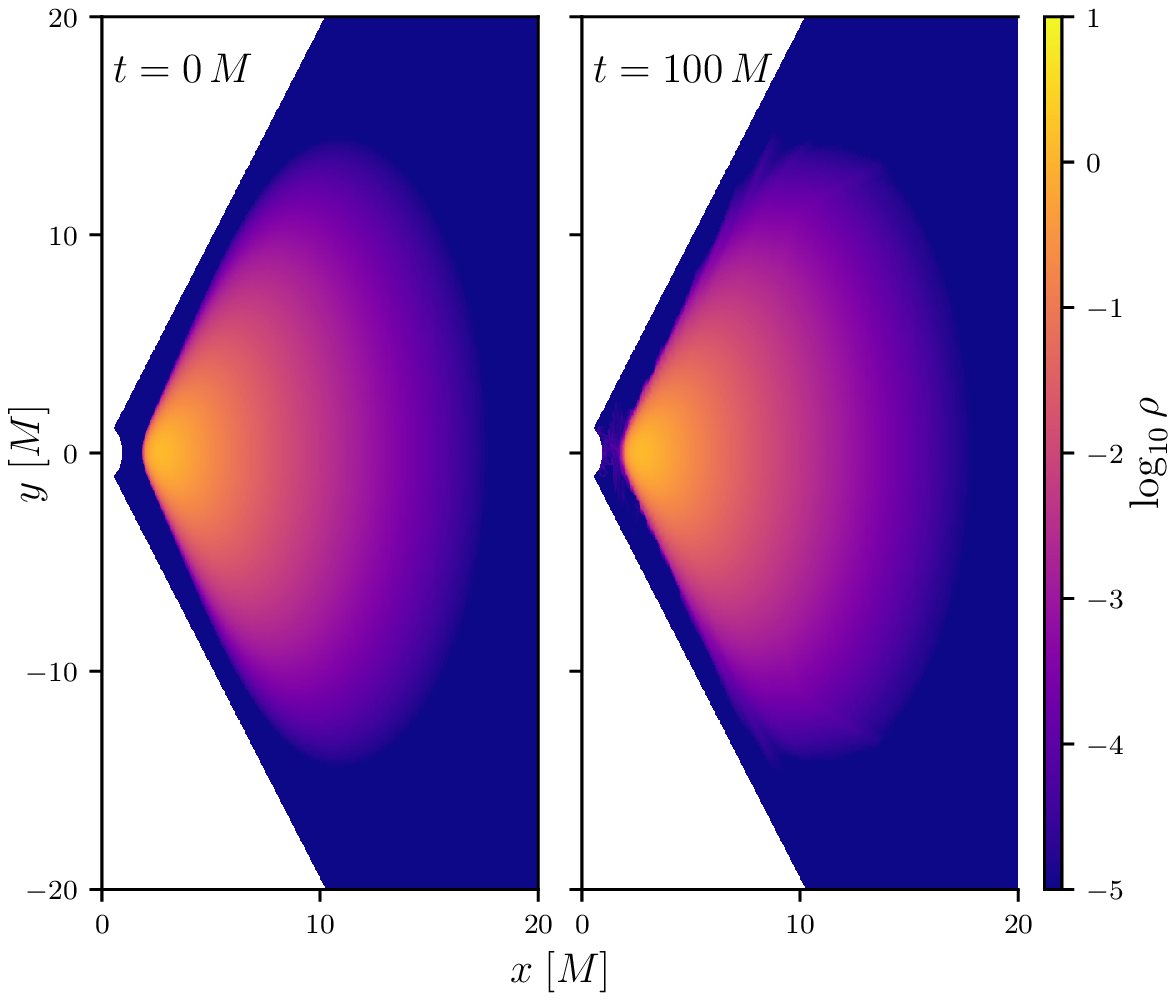}
\includegraphics[width=0.3\linewidth]{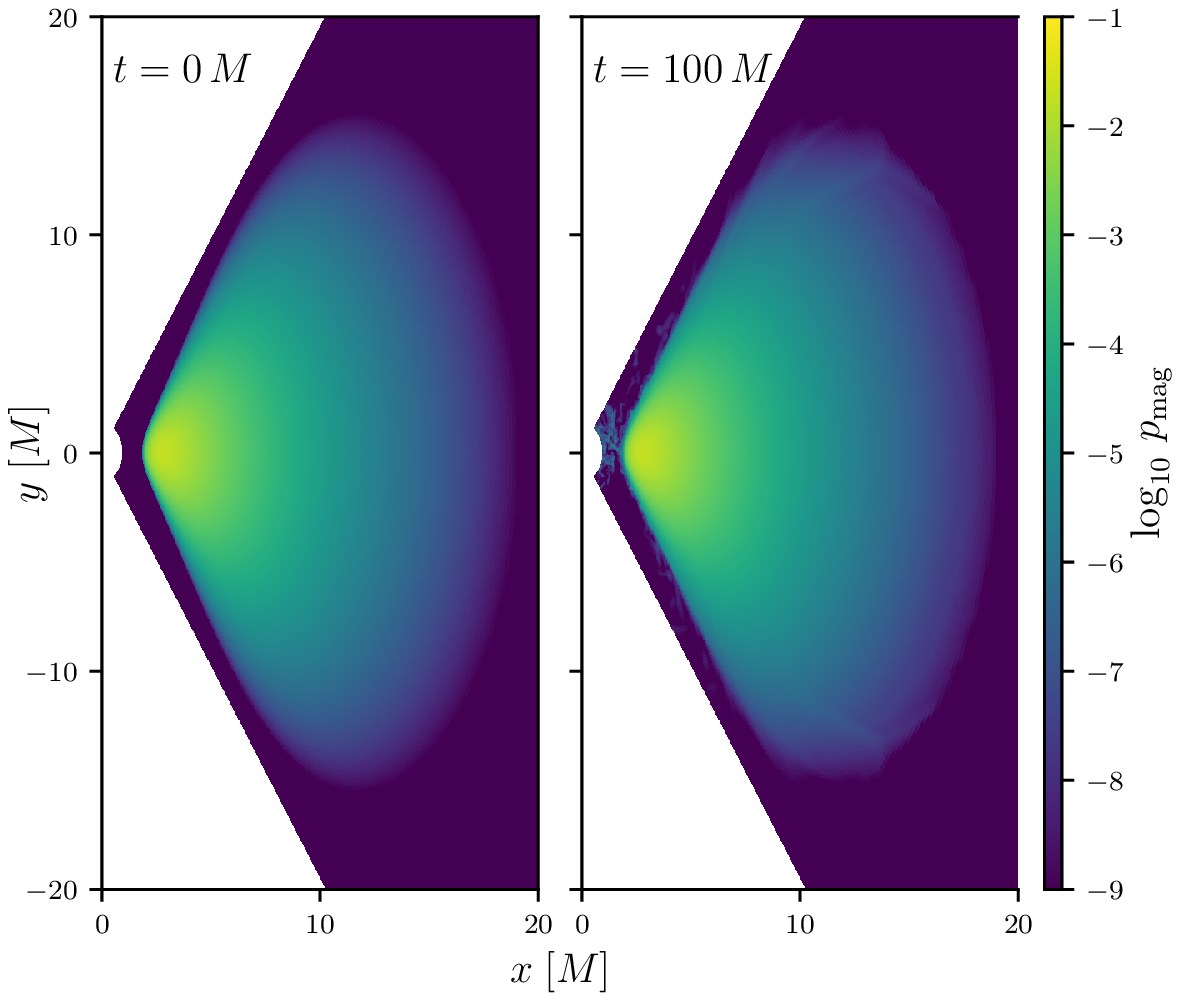}
\includegraphics[width=0.3\linewidth]{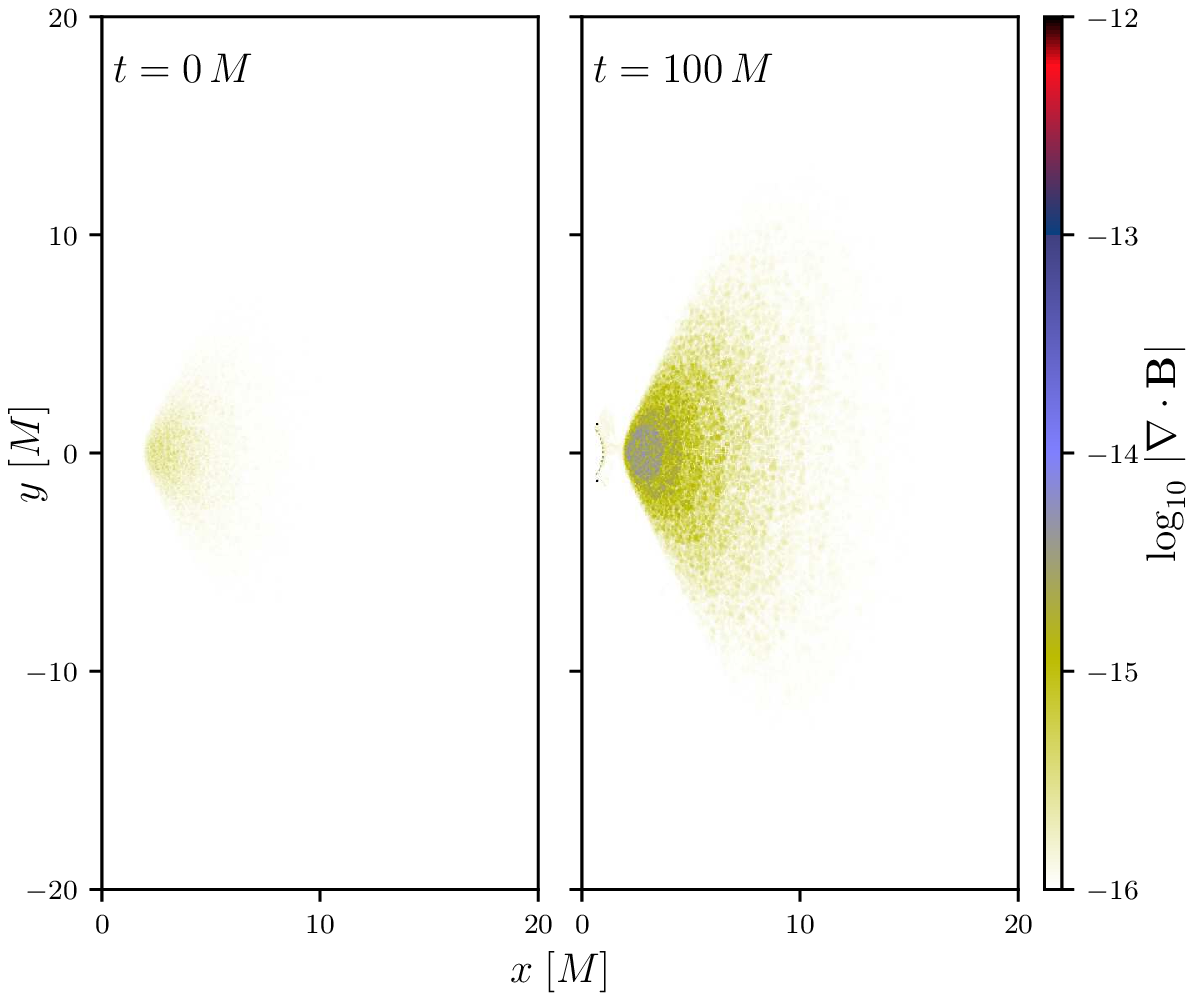}
\caption{Rest mass density, plasma beta and divergence
         of the magnetic field at $t=0$ and $t=100$
         for the run with $N_r=400$ of the
         Komisssarov torus test.}
\label{fig:komissarov-snapshots}
\end{figure*}

\begin{figure}[htbp]
\begin{center}
\includegraphics[width=\linewidth]{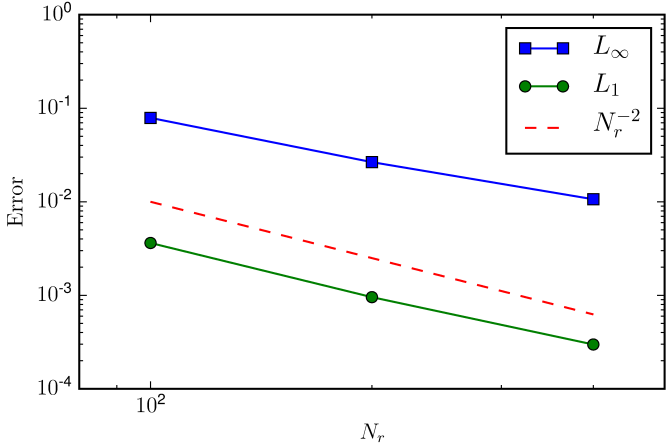}
\caption{$L_\infty$ and $L_1$ norms of the error in density $\rho$ for
  the strongly-magnetised Komissarov torus at $t=70\,M$, \ie after one
  orbital period. The scheme shows second order
  convergence. \label{fig:komissarov-error}}
\end{center}
\end{figure}

Fig.~\ref{fig:komissarov-snapshots} shows the density distribution, the
plasma beta and the divergence of the magnetic field at the initial state
and at $t=100\,M$ for the run with resolution $N_r = 400 $. As can be
seen, the distribution of the displayed quantities is very well preserved
by the scheme.
The $L_1$ and $L_\infty$ errors (calculated by comparing with the last
state with the initial condition) are shown in
Figure~\ref{fig:komissarov-error}. Consistently with the algorithm
employed, second order convergence can be observed.

\section{Magnetized black-hole accretion}
\label{sec:accretion}

In this section we present simulations where the newly implemented
numerical methods are applied to evolve the accretion of plasma onto
black holes, one of the main target applications of the code.

In \citet{Porth2017}, simulations of this kind, performed at uniform
resolution, were used to validate the code by means of qualitative and
quantitative comparisons with the widely used GRMHD code \harm
\citep{Gammie03,Noble2009}. Therefore, this section will be focused in
showing the advantages of AMR and the effects of the choice of divergence
control technique.

For all the simulations in this section, the initial configuration is a
torus in equilibrium around a black hole \citep{Fishbone76}
with spin parameter $a=0.9375$, and thus with the outer event horizon
located at $r=1.348\,M$, where $r$ is the radial Kerr-Schild coordinate.
The inner radius of the torus is at $r_\mathrm{in}=6\,M$, and its density
maximum at $r_\mathrm{max}=12\,M$ (orbital period of $247\,M$ at the
density maximum). The density in the torus is normalized so that it takes
1.0 as its a maximum value. A single-loop poloidal magnetic field is
built from the vector potential $A_{\phi} \propto {\rm max}
(\rho/\rho_{\rm max} - 0.2, 0)$ and is normalized in such a way that the
highest magnetic pressure and the highest thermal pressure are related by
the ratio $\beta=p_\mathrm{fluid,max}/p_\mathrm{mag,max}=100$. To break
this equilibrium state, random perturbations of $4\%$ are added to the
pressure. This eventually triggers the MRI, allowing the plasma to accrete.
It is worth mentioning that even without explicitly adding a perturbation,
numerical errors produced by the discretization would be
amplified to produce a turbulent state very similar to that obtained.
However, in order to have some control on the initial state, the perturbation we
add is significantly larger than numerical errors (which are not easily
predictable) but still small respect to the equilibrium pressure. In addition,
since the initial growth rate of the perturbation depends on its amplitude, the 
saturated state can be reached faster with larger perturbations. For 
this reason, starting with seed perturbations is computationally less
expensive than waiting for the initial discretization errors in the 
equilibrium state to grow. Moreover, the value of $4\%$ is the same used in
\citet{Porth2017}, and is therefore useful for making comparisons.

To avoid vacuum regions, the rest of the simulation is filled with a
tenuous atmosphere with density $\rho_{\mathrm{fl}} = 10^{-4} r^{-3/2}$
and fluid pressure $p_{\mathrm{fl}} = 1/3\times10^{-6}\ r^{-5/2}$. We
reset the density or the pressure whenever they fall below these floor
values. Simulations were evolved using a two-step time-integrator.

As mentioned in Section \ref{sec:coordinates}, several coordinate systems
for black-hole spacetimes are available in \texttt{BHAC}. The simulations
shown here are performed in MKS, as well as in CKS coordinates. Both
coordinate systems possess advantages and weaknesses that must be taken
into account depending on the aspect of the physical system under study,
as will be explained below. The simulations whose results are reported in
this section are summarised in table \ref{tab:accretion_sim}.

\subsection{2D MKS simulations}
\label{sec:accretion2DMKS}

The first pair of simulations is performed in 2D on the meridional plane.
The domain covers $\theta\in [0,\pi]$ and $r \in [1.2,2500]$, and is
resolved using three AMR levels triggered by the L\"ohner scheme, with a
base resolution of $N_r\times N_\theta = 200 \times 100$ and an effective
resolution of $N_r\times N_\theta = 800 \times 400$.
Since the time step for these 2D simulations is not
penalised by the small width of cells in contact with the polar axis, the
parameter $\chm{\vartheta_0}$ of MKS coordinates is set to zero. Reconstruction is
performed using PPM. The purpose of these two simulations is to
investigate the effect of the method utilised for the evolution of the
magnetic field; therefore, the only difference between them was the use
of either the Balsara \& Spicer algorithm or UCT2.

\begin{figure*}
\centering
\includegraphics[width=0.45\linewidth]{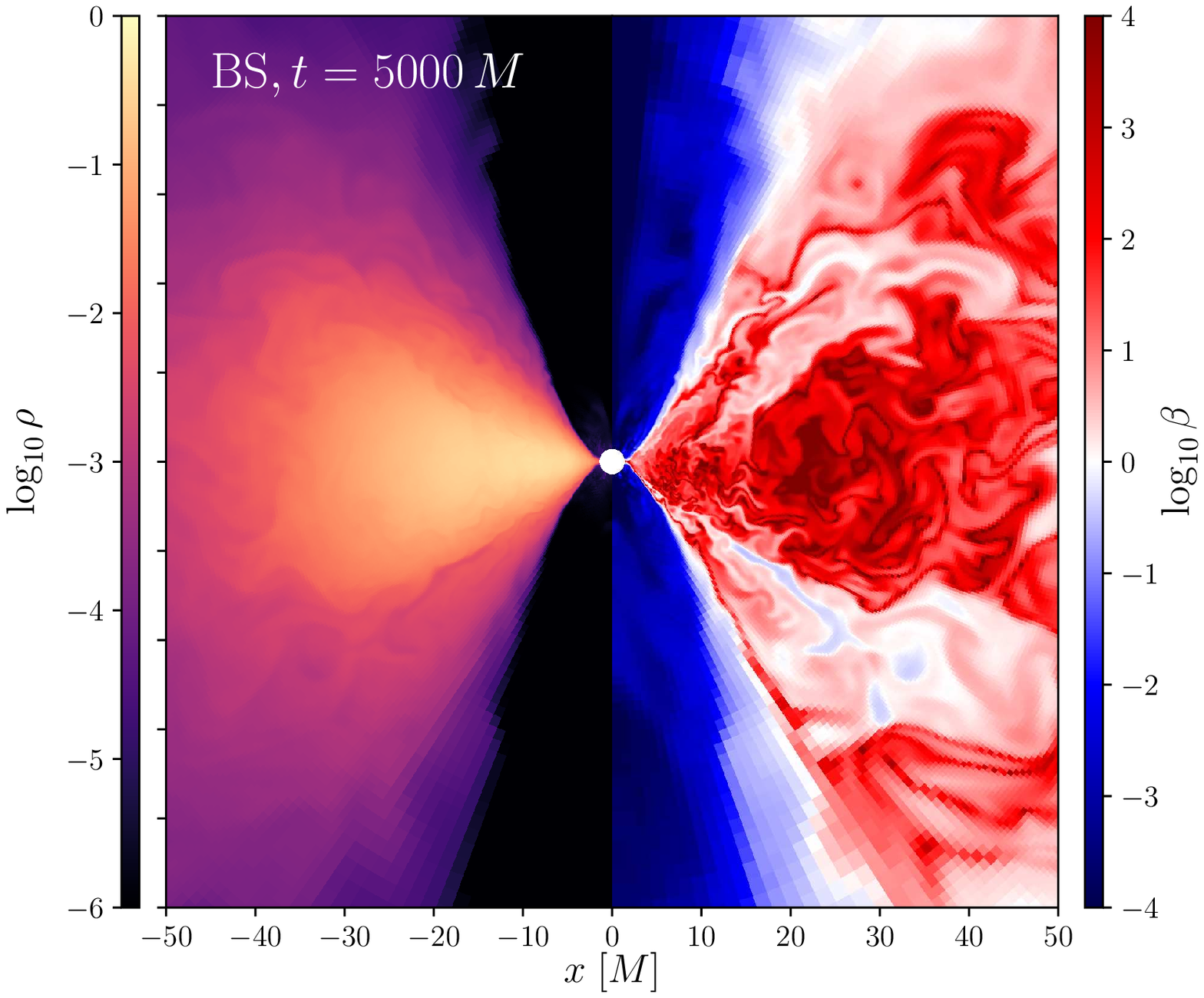}
\includegraphics[width=0.45\linewidth]{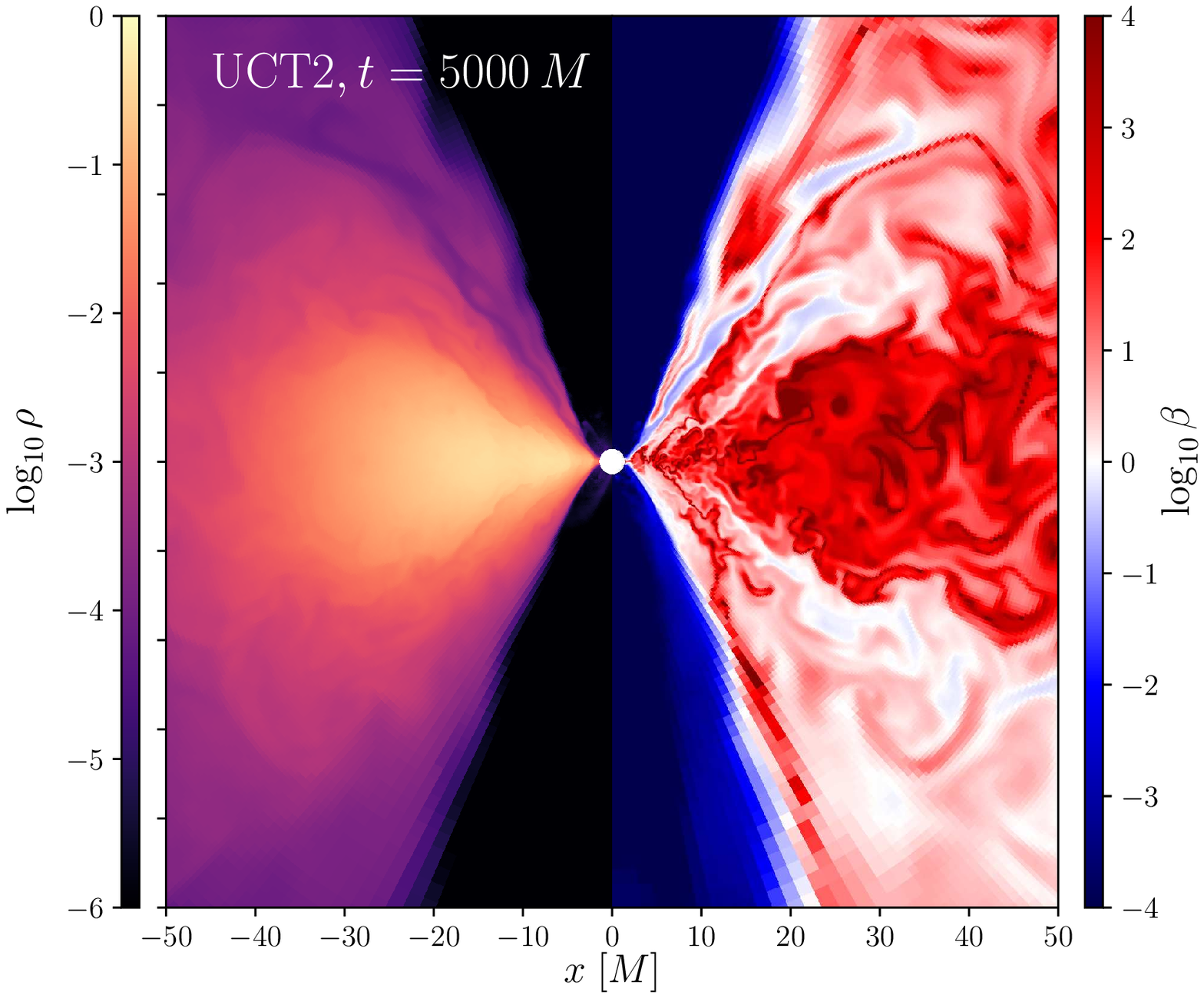}
\caption{Comparison between 2D simulations of black-hole accretion using
  the Balsara \& Spicer method ({\it left}) and the UCT2 algorithm ({\it
    right}) at $t=5000\,M$. For each panel, rest mass density $\rho$ is
  shown at the left and plasma $\beta$ at the right.}
\label{fig:2D_accretion_rhobeta}
\end{figure*}

A comparison at $t=5000$ (Figure \ref{fig:2D_accretion_rhobeta}) displays
what we observe to be systematic differences between the two simulations.
Namely, a higher magnetisation inside the funnel for the UCT2 run and a
different morphology of the jet, which acquires a parabolic shape for the
Balsara \& Spicer run and is more conical for the UCT2 run. Another
difference lays in the transition between the funnel and the mildly
magnetized wind surrounding the jet, which is sharper for the UCT2 case.
\ch{These differences could have important observational consequences,
  for instance, in future high-resolution VLBI images of the SMBH
  candidates in M~87 and Sgr~A*, since radiative models of these sources
  which are able to reproduce the properties of their radio spectrum
  consider that most of the radio emission originates at the funnel wall
  \citep{Moscibrodzka2016}.}

From these simulations, it is possible to see that the choice of method
for evolving the magnetic field can have a visible effect on the flow
behavior. Among these two methods, we would recommend the use of UCT2
due to its upwind properties and to the possibility of using high-order
reconstructions. \ch{Although some asymmetry respect to the equator can
be seen in both the BS and the UCT2 simulations, this is only a consequence
of the chaotic evolution of the initial random perturbations driven by
turbulence, and not of the numerical schemes. Even though not included in the
tests presented in this work, in order to verify that the scheme does not possess 
any directional bias, we have simulated the development of the magnetic 
Kelvin-Helmholtz instability with a controlled perturbation \citep{Toth2002} 
and employing the same methods (PPM and BS or UCT2). When comparing with 
a simulation starting from an initial condition obtained by specular 
reflection, we obtained identical results.}
Finally, it is worth mentioning that the simulation
using the Balsara \& Spicer method was tested previously against another
run performed in an equivalent uniform resolution and using the
widespread flux-CT method. The AMR run was able to achieve a speedup
factor of 7.1, while obtaining quantitatively similar results in the mass
accretion rate and the magnetic flux through the horizon, as discussed in
\citet{Olivares2018a}.

\subsection{3D MKS simulations}
\label{sec:accretion3DMKS}

Since self-sustaining dynamo activity leading to the perpetuation of the
MRI cannot occur in strict azimuthal symmetry
\citep{Cowling33,Balbus1991}, 3D simulations are necessary to study the
accretion flow in the saturated state.

The AMR capabilities of the code become even more important in this case,
where the computational cost of simulations rapidly increases due to the
larger number of computing cells. In addition, the polar coordinate
system naturally leads to a significantly higher resolution close to the
polar axis, which is not always justified, and the narrow cells in these
regions usually cause a penalisation in time-step size.

A possibility to alleviate this limitation is to de-refine cells close to
the polar axis, as done by \citet{White2016}. In typical accretion
scenarios, the polar region is occupied by an evacuated, smooth funnel,
thus no especially high resolution is required. In contrast, equatorial
regions are populated by turbulent structures that need to be properly
resolved.

Newtonian shearing box simulations have shown that an insufficient
resolution can suppress the growth of the MRI when the wavelength of its
fastest growing mode $\lambda_{\rm MRI}$ is resolved with less than six
cells, and move the simulation away from the ratio of magnetic to gas
pressure expected at the saturated state \citep{Sano2004}. A
relativistic version of this criterion has been applied to estimate
whether global accretion simulations are properly resolved \citep[see
  \eg][]{Noble2010,McKinney2012}.

The MRI quality factor is defined as the ratio between the grid spacing
and $\lambda_\mathrm{MRI}$, which approximately gives the number of cells
employed to resolve it. Following \citet{Noble2010}, we define the
relativistic MRI quality factor as
\begin{equation}
\label{eq:lambda_mri}
Q^{(i)}_{\rm MRI} = \frac{\lambda^{(i)}_{\rm MRI}}{ \Delta x^{(i)}}
= \frac{2 \pi }{\Omega} \frac{| b^{(i)} |}{\Delta x^{(i)}\sqrt{\rho h + b^2}} \,,
\end{equation}
where $b^{(i)}$ and $\Delta x^{(i)}$ are, respectively, the magnetic
field and the displacement in the $i$-th direction in the orthonormal
frame co-moving with the fluid, and $\Omega$ is the orbital angular
velocity. Since the value of $b^{(i)}$ depends on
the amplification of the magnetic field experienced during the
simulation, $Q^{(i)}_{\rm MRI}$ can only be determined {\it a posteriori}. For a fluid
rotating in the $\phi$ direction, $Q^{(\theta)}_{\rm MRI}$ is the relevant MRI
quality factor.

The simulation presented in this section, labelled as \texttt{MKS192-UCT}
(see table \ref{tab:accretion_sim}), is performed on a static grid that
has been de-refined at the poles. \ch{More specifically, the AMR blocks
touching both the polar axis and the inner radial boundary, which contain the
smallest cells and therefore determine the global time-step, were forced
to the coarsest refinement level.} In addition to the settings described
at the beginning of this section, the simulation domain spans $ r = [1.18
  \, M , 10000/3\, M ]$ and the extent in $\theta$ and $\phi$ subtends
the whole $4\pi$ steradians solid angle. The resolution at the base level
is $N_r \times N_\theta \times N_\phi = 128 \times 48 \times 48 $, and
the simulation contains 3 AMR-levels, giving an effective resolution of
$512 \times 192 \times 192 $ cells. As an additional measure to prevent
the time-step penalisation from the poles, this time the MKS
$\chm{\vartheta_0}$-parameter is set to 0.25.

\begin{figure*}
\centering
\includegraphics[width=0.44\linewidth]{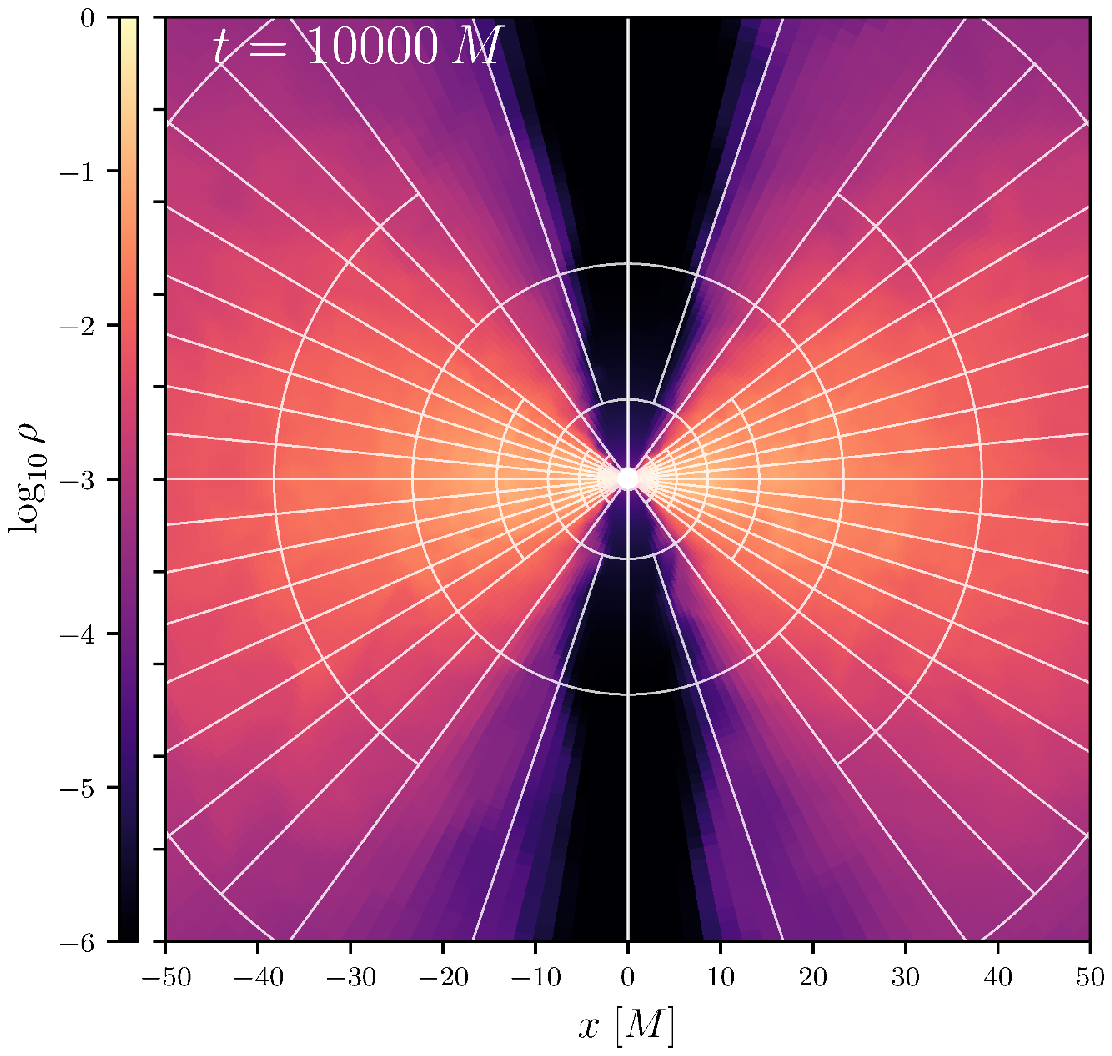}
\includegraphics[width=0.44\linewidth]{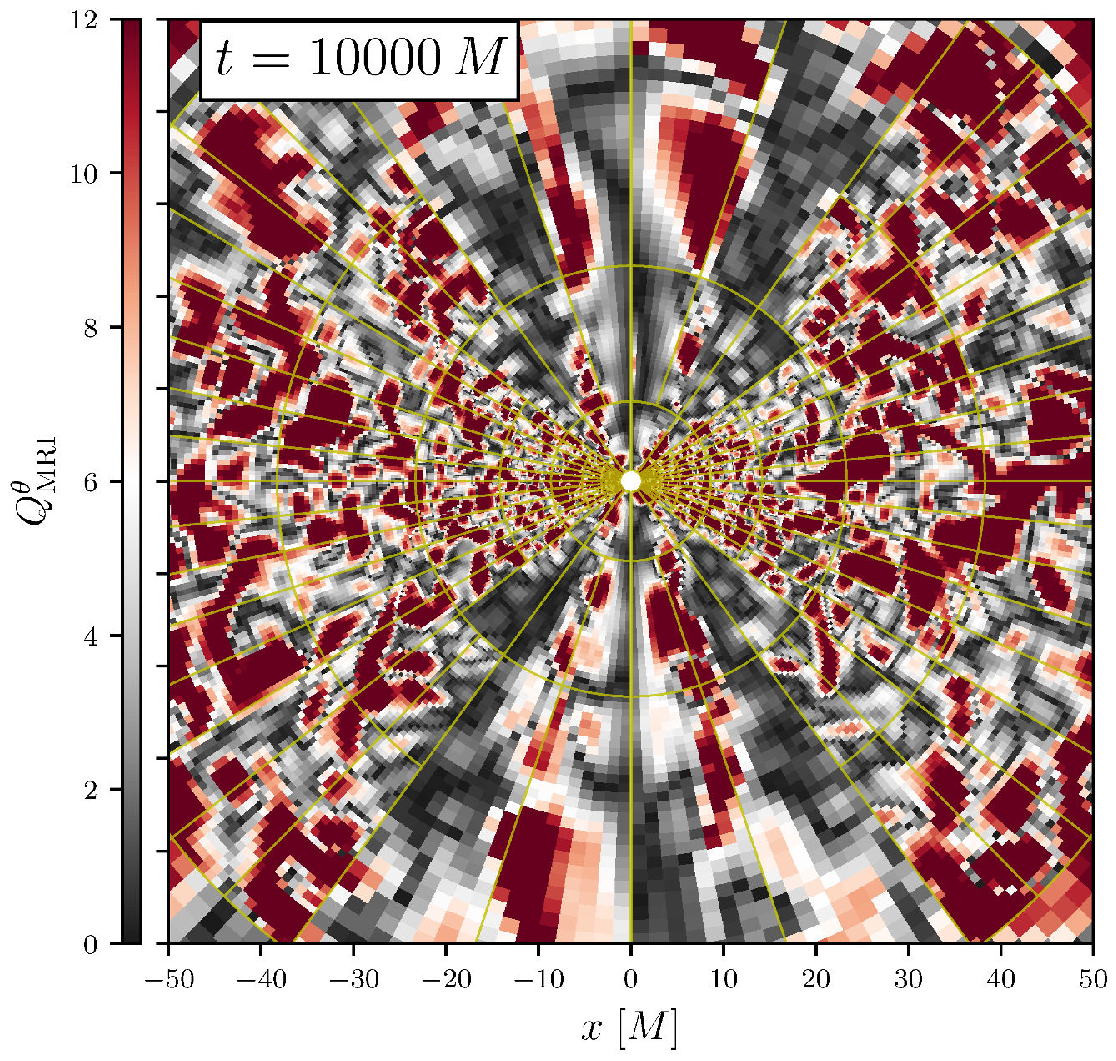}
\caption{Logarithmic rest-frame density ({\it left panel}) and MRI
  quality factor ({\it right panel}) for simulation \texttt{MKS192-UCT},
  showing the AMR blocks of $N_r\times N_\theta \times N_\phi =
  32\times 8 \times 8$ cells each.
  The grid is able to concentrate resolution in
  the disk in order to achieve $Q^{(\theta)}_{\rm MRI}>6$ \ch{in the torus},
  as recommended in \citet{Sano2004}, while saving computational
  costs by de-refining the polar regions.}
\label{fig:toruscc192rhoq}
\end{figure*}

Figure \ref{fig:toruscc192rhoq} shows a snapshot of simulation
\texttt{MKS192-UCT} at $t= 10\,000\,M$. The AMR blocks, of $32 \times 8
\times 8$ cells each, are drawn in white lines. The left panel of that
figure shows the logarithm of density, while the right panel shows the
MRI-quality factor. It can be observed that the simulation grid is able
to maintain $Q^{(\theta)}_{\rm MRI}>6$ in the disk without \ch{requiring
a similarly high resolution at the polar regions, which are occupied by
a smooth low-density outflow and where MRI-driven angular momentum
transport does not play an important role}. It is necessary to keep in
mind that places where $Q^{(\theta)}_{\rm MRI} \sim 0$ within the disc
correspond to regions where the magnetic-field changes sign
(see Eq. \ref{eq:lambda_mri}).

\begin{figure}
\centering
\includegraphics[width=\linewidth]{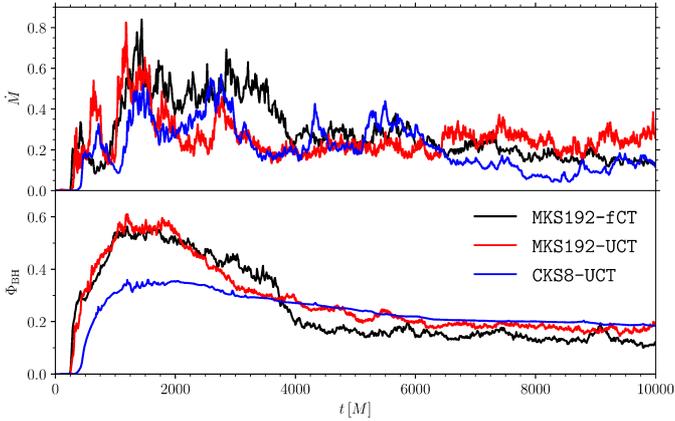}
\caption{Mass-accretion rate ({\it top}) and magnetic flux through the
  horizon ({\it bottom}) for the 3D simulations shown in this work in MKS
  and CKS coordinates using UCT2 (see table \ref{tab:accretion_sim}), and
  for the highest-resolution simulation in MKS coordinates shown in
  \citet{Porth2017}, using flux-CT.}
\label{fig:accretionfluxes}
\end{figure}

As can be seen in Figure \ref{fig:accretionfluxes}, the mass-accretion
rate $\dot{M}$ and magnetic flux through the horizon $\Phi_{\rm BH}$
\citep{Tchekhovskoy2011} are consistent with those from
the highest resolution of \citet{Porth2017} (labelled here as
\texttt{MKS192-fCT}, see table \ref{tab:accretion_sim}), which used the
same initial condition and the numerical methods validated in that work
including the flux-CT scheme. In both cases, the behavior consists of a
transient growth ($t <~ 5000 \,M$) followed by a quasi-stationary state
after the saturation of MRI ($t >~ 5000 \,M$).

As an additional comparison tool, we compute time- and disk-averaged
profiles similar to those shown in
\citet{Beckwith2008,Shiokawa2012,White2016,Porth2017}. For a quantity
$q(t,x^i)$, these averages are defined as
\begin{equation}
\langle q \rangle = \frac{\int_{t_{\rm min}}^{t_{\rm max}}
                          \int_{\theta_{\rm min}}^{\theta_{\rm max}}
                          \int_{0}^{2 \pi} \sqrt{-g}\ q(t,r,\phi,\theta)\ d\phi\ d\theta\ dt}
                         {\int_{t_{\rm min}}^{t_{\rm max}}
                          \int_{\theta_{\rm min}}^{\theta_{\rm max}}
                          \int_{0}^{2 \pi} \sqrt{-g}\ d\phi\ d\theta\ dt} \,,
\end{equation}
were $\theta_{\rm min} = \pi/3$, $\theta_{\rm min} = 2 \pi/3$,
$t_{\rm min} = 5000\,M$ and $t_{\rm min} = 10000\,M$.

As shown in Figure \ref{fig:accretionprofiles}, the averaged profiles of
\texttt{MKS192-UCT} show a reasonable quantitative agreement with those
of \texttt{MKS192-fCT}, \ch{despite the use of different methods and 
resolution. It should be noted, however, that the angular resolution in theta,
which is essential to capture the MRI, is effectively the same for the two
simulations performed in MKS coordinates, due to the use of three
AMR levels in the former. As will be shown in Section \ref{sec:accretion3DCKS},
a better agreement is found with simulation \texttt{CKS8-UCT},
which employs CKS coordinates,
with differences arising from well understood reasons. The fact that such a better
agreement can be found between simulations using different coordinate systems,
and thus completely different spatial discretizations, shows that the
differences between \texttt{MKS192-fCT} and \texttt{MKS192-UCT} are likely
more a consequence of the method employed to evolve the magnetic field
rather than of the slightly different radial resolution.}

\begin{table*}
\centering
\begin{tabular}{l l l c l l}
\hline
Simulation  & Coordinates & Domain $[M]$ & AMR levels & Base resolution & $\boldsymbol{B}$-evolution \\
\hline
\texttt{2DMKS-BS}  & MKS & $[1.2,2500]$ & 3 & $200\times 100$  & BS \\
\texttt{2DMKS-UCT} & MKS & $[1.2,2500]$ & 3 & $200\times 100$  & UCT2 \\
\texttt{MKS192-fCT} & MKS & $[1.2,2500]$ & 1 & $384 \times 192 \times 192$  & flux-CT \\
\texttt{MKS192-UCT} & MKS, $\chm{\vartheta_0}=0.25$ & $[1.2,1000/3]$ & 3 & $128\times 48 \times 48$  & UCT2 \\
\texttt{CKS8-UCT} & CKS & $x,y\in[-500,500]$ & 8 & $96\times 96 \times 192$  & UCT2 \\
                  &     & $z\in[-1000,1000]$ &   &                          &      \\
\hline
\end{tabular}
\caption{{\bf A summary of the simulations referred to in Section
    \ref{sec:accretion}.}  When not specified, the MKS parameters $R_0$
  and $\chm{\vartheta_0}$ are zero. The base grid resolution is displayed as $N_r \times
  N_\theta (\times N_\phi) $ for the MKS simulations and as $N_x \times
  N_y \times N_z $ for the CKS.}
         \label{tab:accretion_sim}
\end{table*}

\subsection{3D CKS simulations}
\label{sec:accretion3DCKS}

Despite the strategies to avoid slow time steps from polar regions
described in Section \ref{sec:accretion3DMKS} and the self-consistent
treatment of poles by rearranging the grid topology described in Section
\ref{sec:poles}, the poles of a coordinate system still represent an
inhomogeneity in the numerical domain that could in principle introduce
artefacts or a directional bias in the simulation.

In addition, although logarithmic grids in polar coordinates ensure
machine-precision conservation of angular momentum\footnote{
\ch{By adding to one cell the numerical flux that was subtracted from its
neighbors (see Section \ref{sec:finite-volume}), finite-volume methods
achieve machine-precision conservation of the {\it conserved}
variables \citep[see \eg][]{Leveque2002,Rezzolla_book:2013}.
When solving the GRMHD equations in polar coordinates,
these include the covariant momentum in the $\phi$-direction.}}
and are very efficient
at resolving small-scale features when the interesting dynamics occur
near the origin, they loose resolution far away from it.

Therefore, in order to study systems that are extended in space and for
which an artificial directional bias could be misinterpreted as a
physical property of the system, it might be useful to resort to
coordinate systems that are more spatially homogeneous. An obvious
choice are coordinate systems based on Cartesian coordinates, which are
often used for GRMHD simulations in full general relativity \citep[see
  \eg][]{Giacomazzo:2007ti,Etienne2015}. However, the necessity to
properly resolve the black-hole horizon and the MRI renders large scale
($\sim 10^3\ r_g$) simulations impossible, unless some form of mesh
refinement is used.

In this section we present a simulation performed in CKS coordinates,
labelled as \texttt{CKS8-UCT} in Table \ref{tab:accretion_sim}. A
combination of AMR and static mesh refinement is used to ensure
sufficient resolution at the black-hole horizon and the MRI
turbulence while following the jet with dynamic mesh refinement.

\begin{figure*}
\centering
\includegraphics[width=\linewidth]{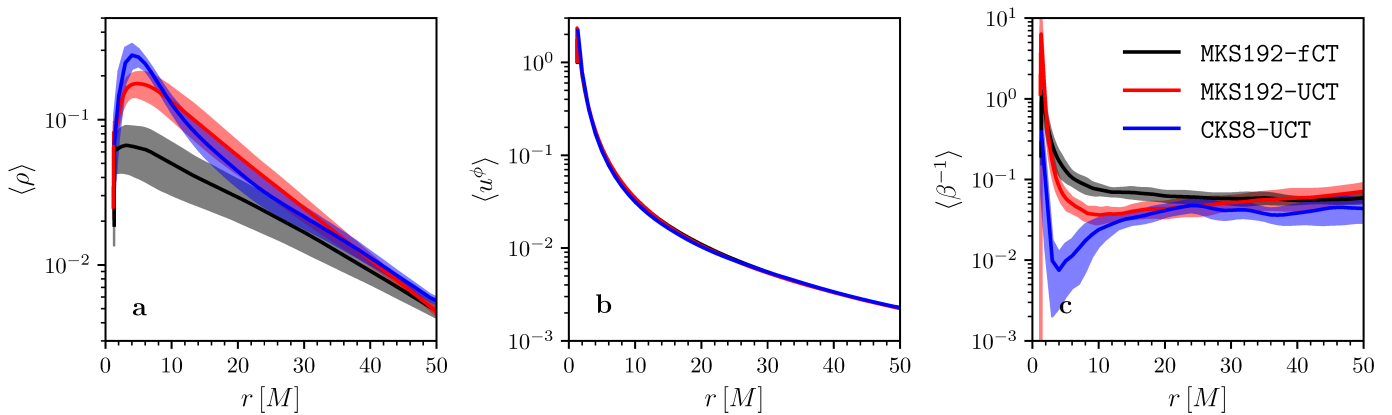}
\caption{Disk- and time-averaged profiles of rest mass density
  (\textit{a}), $\phi$-component of the 4-velocity ({\it b}), and inverse
  plasma $\beta$ ({\it c}), in the interval 5000-10000 $M$, for the 3D
  simulations of magnetized accretion referred to in this work (see table
  \ref{tab:accretion_sim}). Shaded regions show the standard deviation
  from the average value for each simulation. \ch{(These are, however,
    hardly noticeable for $u^\phi$ due the small standard deviation and
    the excellent agreement between the three curves.)} After taking into
  account the different numerical methods and coordinates employed, the
  three simulations show reasonably consistent profiles, with the
  deviations in density and plasma $\beta$ at $r_{KS} < 10\ M$ for
  simulation \texttt{CKS8-UCT} probably caused by a poor resolution of
  the MRI in that region (see also Figure \ref{fig:cartesian0800}).}
\label{fig:accretionprofiles}
\end{figure*}

The domain is structured as a set of nested boxes for which a different
maximum refinement level is specified. The highest AMR level can be
reached only by the innermost box, which contains the black-hole horizon.
In order to follow the jet dynamics, refinement is triggered by
variations in the plasma magnetisation $\sigma = b^2/\rho$ and the
density $\rho$, using the L\"ohner scheme. In order to limit the overhead
by alternating refinement and de-refinement of the same regions,
re-gridding is performed only every 1000 iterations (since hierarchical
time-stepping has not yet been implemented in \texttt{BHAC}, these
correspond to the time-step of the finest grid). The simulation employs 8
AMR levels, with a base resolution of $N_x\times N_y\times N_z = 96
\times 96 \times 192$, and extending over $x,y \in [-500\,M,500\,M]$, and
$z \in [-1000\,M,1000\,M]$. The magnetic field is evolved using the UCT2
algorithm. In order to prevent an unphysical outflow from the black-hole
interior, we apply cut-off values for the density and pressure in the
region $r<0.5(r_{\rm H- } + r_{\rm H+ })$, where $r_{\rm H- }$ and
$r_{\rm H+ }$ are the location of the inner and outer event horizons. In
particular, we set $\rho_{\rm cut} = 10^{-2}$ and $p_{\rm cut} = 10^{-4}$.

\begin{figure*}
\centering
\includegraphics[width=\linewidth]{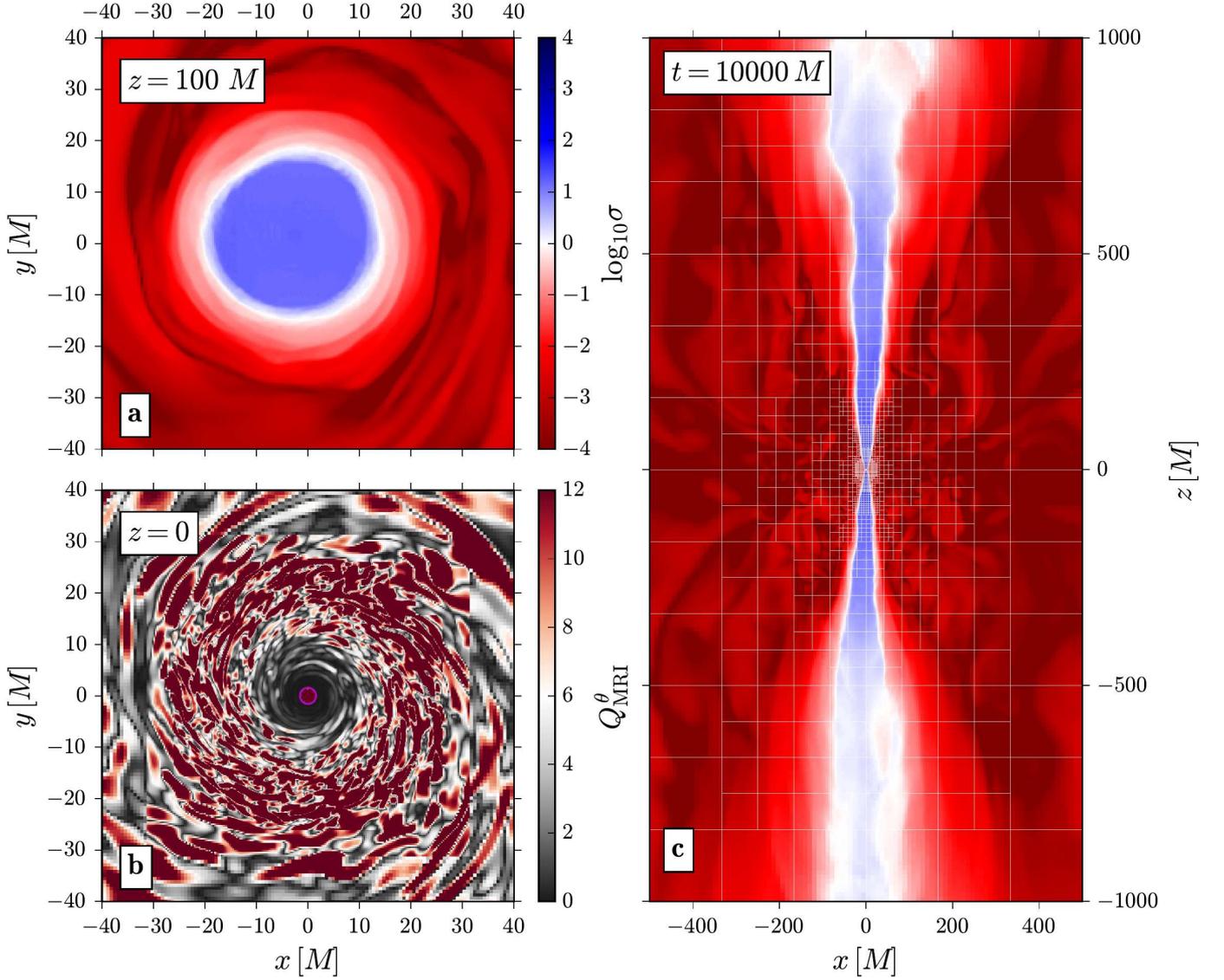}
\caption{Cuts of the simulation \texttt{CKS8-UCT} (see Table
  \ref{tab:accretion_sim}) at $t=10000\ M$, showing logarithmic plasma
  magnetisation $\sigma = b^2/\rho$ and MRI quality factor in the
  $\theta$-direction, $Q^{(\theta)}_{\rm MRI}$. {\it Panel a}: Cross
  section of the jet at $z=100\ M$. {\it Panel b}: MRI-quality factors
  in the disk at $z=0$. A magenta circumference marks the black-hole
  outer horizon. {\it Panel c}: AMR blocks, of $16^3$ cells, following
  the propagation of the jet.}
\label{fig:cartesian0800}
\end{figure*}

Figure \ref{fig:cartesian0800} displays different cuts of the simulation
at $t=10000\ M$. The two left panels are horizontal cuts at $z=100\ M$
(panel {\it a}) and $z=0$ (panel {\it b}), and the right panel is a
vertical cut at $y=0$ (panel {\it c}). In panel ({\it a}) it is possible
to appreciate a cross-section of the jet, which is now completely free to
wobble and to change shape independently of the coordinate surfaces.
Panel ({\it c}) shows the automated mesh refinement following the
evolution of the jet, indicated by a high plasma magnetisation
$\sigma=b^2/\rho$, as it propagates in the $\pm z$ direction. Also in
this panel it can be noticed that the shape of the jet does not appear
constrained by the coordinate surfaces. In a future work we will provide
a more detailed comparison of the jet dynamics in simulations using
Cartesian and spherical grids.

At the same time as the jet is resolved with such detail, panel ({\it b})
shows that in most of the disk MRI is resolved with high quality factors,
being the only exception the region for which $r_{\rm KS} < 10 \ M$.

To quantitatively compare the results of this Cartesian simulation with
those presented in the previous sections, also in this case we compute
the mass-accretion rate and magnetic flux through the horizon as well as
time- and disk-averaged profiles in the same intervals as mentioned in
Section \ref{sec:accretion3DMKS}.

In Figure \ref{fig:accretionfluxes}, the mass-accretion rate shows a
remarkably consistent behavior for all of the three simulations, with
variations being a consequence of the turbulent nature of the process.
On the other hand, the absolute magnetic flux through the horizon, though
overall consistent in magnitude, shows significantly less variations for the
Cartesian case. This is accompanied by a smaller maximum in the initial
transient growth. This could likely be attributed to a poorer resolution
of the disk region which hinders magnetic-field amplification due to MRI.
In fact, although overall high quality factors are obtained within the
disk, the decrease in $\lambda^{(\theta)}_{\rm MRI}$ due to the density
increase close to the black hole is not accompanied by a corresponding
increase in resolution as is the case for spherical polar coordinates
(\cf panel {\it b} of Figure \ref{fig:cartesian0800} and right panel of
Figure \ref{fig:toruscc192rhoq}). A more adequate resolution could be
achieved by allowing higher AMR levels in this region.

The averaged profiles shown in Figure \ref{fig:accretionprofiles} are as
well in good agreement with the other 3D simulations, especially with
\texttt{MKS192-UCT}, which employs the same algorithm to evolve the
magnetic field. The agreement is practically perfect for the
$\phi$-component of the 4-velocity in all three simulations and it
remains mostly within one standard deviation between simulations
\texttt{CKS8-UCT} and \texttt{MKS192-UCT}. The most important deviations
in density and plasma $\beta$ at $r_{\rm KS} < 10\ M$ with respect to
\texttt{MKS192-UCT} could likely be attributed to the poorer resolution
of MRI in that region, which hinders the amplification of the magnetic
field and the angular momentum transport leading to an accumulation of
mass in that region. In fact, panel ({\it c}) of Figure
\ref{fig:accretionprofiles} is consistent with a result from
\citet{Sano2004}, namely that a poor resolution leads to values of the
magnetic pressure (and thus of $\beta^{-1}$) smaller than those expected
at the saturation of MRI.

Finally, \ch{Table \ref{tab:accretion_resolution}} lists several properties of runs
\texttt{CKS8-UCT} and \texttt{MKS192-UCT} at the end of the simulation,
related to how well they can resolve physics.
These are: number of cells resolving the
horizon, total cell population, time-step and average volume of occupied
cells, \ie the average proper volume of cells which contain matter coming
from the disk, which is identified by a passively advected tracer.
Unsurprisingly, it can be seen that, although MKS simulations are able to
resolve better the horizon and reach higher MRI quality factors in the
disk region, overall the domain is more resolved in the CKS case. As a
consequence, simulations in Cartesian coordinates could be useful to
study the large-scale effects on the jet produced by finer features
arising from Kelvin Helmholtz or kink instabilities, as those visible
in Figure \ref{fig:cartesian0800}.

Furthermore the small sizes of cells close to the polar axis or the outer
event horizon produces a penalisation in time-step for MKS simulations
which is absent for the Cartesian case, which is thus able to advance
more physical time per iteration. However, the price to pay for a more
resolved domain in the latter is a much larger cell population which
currently is updated simultaneously, significantly increasing the
computational cost of these simulations. In the future, this limitation
will be overcome to some extent by adopting a hierarchical time step as
is done in several AMR-codes \citep[see
  \eg][]{Cunninghametal09,Liska2018}.

In summary, due to the advantages mentioned above, Cartesian adaptive
meshes appear as a very interesting resource to study large-scale jet
propagation in simulations which self consistently include the jet
engine, as well as other systems for which no symmetry is to be assumed
{\it a priori}, as tidal disruption events or precessing disks.
Currently, the Cartesian simulation described above is being used to
model the appearance of the jet-launching region in M~87
\citep{Davelaar2019}.

\begin{table}
\centering
\begin{tabular}{l r r r r}
\hline
Simulation & $N_{\rm H}$ & 
 $\langle \Delta V \rangle_{\rm occ}^\frac{1}{3}$ & $N_{\rm cells}$ & $\Delta t$ \\
\hline
 \texttt{CKS8-UCT}     &  $2\,880$  
               & 2.9               &  70 811 648               & $2.44\times 10^{-2}$ \\
 \texttt{MKS192-UCT}   &  $20\,736$ 
                   & 7.6               &  6 144 000           & $4.71\times 10^{-3}$ \\
\hline
\end{tabular}
\caption{{\bf Comparison between Cartesian and spherical runs for the
  same accretion problem.}  The quantities shown are: number of cells
  resolving the outer event horizon $N_{\rm H}$,
  cubic root of the average proper volume per occupied cell,
  total cell population and time step, with all quantities
  measured at time $t=10000\ M$. \label{tab:accretion_resolution}}
\end{table}

\section{Conclusion}
\label{sec:conclusion}

We described in detail new additions to the GRMHD code \texttt{BHAC},
namely three CT algorithms based on staggered meshes, as well as AMR
strategies compatible with them. The variants of CT implemented in
\texttt{BHAC} are the arithmetic average of \citet{Balsara99} (BS) and
the two upwind schemes by \citet{londrillo2004divergence} and
\citet{DelZanna2007} (UCT), which allow the use of high-order
reconstructions.

In order for the divergence of the magnetic field to be zero to machine
precision across coarse/fine boundaries and during the creation and
destruction of blocks caused by AMR, special divergence-preserving
restriction and prolongation operators are required. We derived and
employ a class of such operators that generalises those obtained for
Cartesian geometry by \citet{Toth2002}.
In addition, we presented technical information on the data structures
used, the ghost-cell exchange which was re-designed for staggered
variables, and the treatment of the poles in spherical and cylindrical
coordinates, as well as on divergence-preserving boundary conditions.

We validated these new additions by showing the results of tests commonly
used in the community, specifically, \citet{Gardiner2005}'s loop
advection test and the cylindrical explosion from \citet{Komissarov1999}
in special relativity, and the magnetised stationary torus from
\citet{Komissarov2006a} in general relativity. We observed that,
in agreement with analogous tests present in the literature,
UCT methods generate less spurious oscillations than the flux-CT and BS
methods, both still widely used in GRMHD and MHD codes, and that the
algorithm converges at the expected rate.

By performing 2- and 3-dimensional simulations of magnetized accretion
onto a black hole, we could notice that the BS method can give different
results in physically relevant problems as compared to UCT methods. Due
to its upwind properties and to the possibility of employing high-order
reconstructions in UCT methods, we decided to use the latter schemes for
future simulations.
In addition, we showed that the code's AMR capabilities can be exploited in black
hole accretion simulations in order to eliminate the penalisation in
time-step caused by the small width of the cells around the pole in
spherical coordinates, while maintaining an optimal resolution at the
turbulent equatorial regions.

As an example of a problem inaccessible, or at least extremely expensive
for non-AMR codes, which can be performed using the new capabilities
added to the \texttt{BHAC} infrastructure, we simulated black-hole
accretion in Cartesian Kerr-Schild coordinates. The use of a Cartesian
mesh could permit the study of jet dynamics including self consistently
the black-hole engine, while avoiding any possible directional bias
introduced in the mesh by the presence of a polar axis. Furthermore, AMR
can be used to accurately simulate magnetohydrodynamical instabilities
between the disk wind and the jet occurring during its propagation,
making similar set-ups extremely useful for self-consistently modelling
sources such as M~87 \citep{Davelaar2019} and Cen A \citep{Fromm2019},
as will be shown in future work. As mentioned before, Cartesian
coordinates and AMR could be very useful also in accretion scenarios
without symmetries, such as tidal disruption events or precessing disks.

Recent developments in the code that will be presented as well in
forthcoming publications include an accurate modelling of electron
thermodynamics \citep{Mizuno2019} and the addition of a module for resistive
GRMHD \citep{Ripperda2019} which also employs the staggered grid
infrastructure developed in the present work.
A set of comprehensive tests showing consistency between results obtained
with several state-of-the-art GRMHD codes including \texttt{BHAC} for the same
accretion problem will be  presented in \citet{PorthChatterjeeEtAl2019}.

All of these tools, together with the current capabilities of
\texttt{BHAC}, are meant to contribute to a detailed modelling of
strong-field field phenomena in astrophysics, which is becoming
increasingly relevant for international efforts such as the EHT and
BlackHoleCam \citep{Akiyama2019_L1, Akiyama2019_L2, Akiyama2019_L3,
  Akiyama2019_L4, Akiyama2019_L5,Akiyama2019_L6}.

\section*{Acknowledgements}
We would like to thank Alejandro Cruz Osorio, Lukas Weih,
David Kling, Jonas K\"ohler and Mariafelicia de Laurentis for useful
discussions. This research is supported by the ERC synergy grant
"BlackHoleCam: Imaging the Event Horizon of Black Holes" (Grant
No. 610058), by ``NewCompStar'', COST Action MP1304, by the LOEWE-Program
in HIC for FAIR, and by the European Union's Horizon 2020 Research and
Innovation Programme (Grant 671698) (call FETHPC-1-2014, project
ExaHyPE). During the completion of this work, HO was supported in part by
a CONACYT-DAAD scholarship. \ch{The simulations were performed in part on
the SuperMUC cluster at the LRZ in Garching, the LOEWE cluster at the CSC in
Frankfurt, the Iboga cluster at the ITP Frankfurt, the HazelHen cluster
at the HLRS in Stuttgart, as well as on the Dutch National Supercomputing
cluster Cartesius, funded by the NWO computing grant 164.} We
acknowledge technical support from Thomas Coelho.


\end{document}